\documentclass[graybox,vecarrow]{svmult}

\newcommand {\bea}{\begin{eqnarray}}
\newcommand {\eea}{\end{eqnarray}}
\newcommand {\be}{\begin{equation}}
\newcommand {\ee}{\end{equation}}
\newcommand {\bc}{\begin{center}}
\newcommand {\ec}{\end{center}}

\usepackage{mathptmx}       
\usepackage{helvet}         
\usepackage{courier}        
\usepackage{type1cm}        
\usepackage{makeidx}         
\usepackage{graphicx}        
\usepackage{multicol}        
\usepackage[bottom]{footmisc}

\begin{document}

\title*{Scaling Flows and Dissipation in the Dilute Fermi Gas
at Unitarity}

\author{T.~Sch\"afer and C.~Chafin}

\institute{Department of Physics, North Carolina State University,
Raleigh, NC 27695}

\maketitle

\abstract{We describe recent attempts to extract the shear 
viscosity of the dilute Fermi gas at unitarity from experiments
involving scaling flows. A scaling flow is a solution of the 
hydrodynamic equations that preserves the shape of the density 
distribution. The scaling flows that have been explored in the 
laboratory are the transverse expansion from a deformed trap 
(``elliptic flow''), the expansion from a rotating trap, and 
collective oscillations. We discuss advantages and disadvantages 
of the different experiments, and point to improvements of the 
theoretical analysis that are needed in order to achieve 
definitive results. A conservative bound based on the current 
data is that the minimum of the shear viscosity to entropy 
density ration is $\eta/s\leq 0.5\,\hbar/k_B$. }

\section{Introduction}
\label{sec_intro}

 A cold, dilute Fermi gas of non-relativistic spin 1/2 particles
interacting via a short range interaction tuned to infinite 
scattering length, commonly referred to as the unitary Fermi gas,
provides a new paradigm for many strongly correlated quantum systems 
\cite{Bloch:2007,Giorgini:2008}. In this contribution we focus on 
non-equilibrium aspects of the unitary Fermi gas, in particular 
its shear viscosity \cite{Schafer:2009dj}. The shear viscosity 
of a liquid composed of weakly coupled quasi-particles can be 
estimated as 
\be
\label{eta_mfp}
\eta = \frac{1}{3}\,n p l_{\it mfp}\, ,
\ee
where $n$ is the density, $p$ is the average momentum of the
particles, and $l_{\it mfp}$ is the mean free path. The mean free
path can be written as $l_{\it mfp}=1/(n\sigma)$ where $\sigma$
is the transport cross section. Equ.~(\ref{eta_mfp}) implies that 
the shear viscosity decreases as the strength of the interaction 
increases. In the unitary gas the cross section saturates the $s$-wave 
unitarity bound $\sigma=4\pi/k^2$, where $k$ is the scattering momentum, 
and we expect the shear viscosity to be unusually small. 

 Danielewicz and Gyulassy pointed out that the Heisenberg uncertainty 
relation imposes a bound on the product of the average momentum and 
the mean free path, $p l_{\it mfp}\geq \hbar$, and concluded that 
$\eta/n \geq \hbar$ \cite{Danielewicz:1984ww}. This is not a precise 
statement: The kinetic estimate in equ.~(\ref{eta_mfp}) is not valid 
if the mean free path is on the order of the mean momentum. A more 
precise bound has recently emerged from holographic dualities in 
string theory. In this context the natural quantity to consider is 
not the ratio $\eta/n$, but $\eta/s$, where $s$ is the entropy density. 
Policastro, Son and Starinets showed that in ${\cal N}=4$ supersymmetric 
QCD the strong coupling limit of $\eta/s$ is equal to $\hbar/(4\pi k_B)$ 
\cite{Policastro:2001yc}. It was later shown that the strong 
coupling limit is universal in a large class of field theories,
and it was conjectured that $\eta/s \geq\hbar/(4\pi k_B)$ is a 
general lower bound, valid for all fluids \cite{Kovtun:2004de}.

 Are there any fluids in nature that attain or possibly violate 
the proposed bound? A fluid that saturates the bound has to be a 
quantum fluid (because $\eta$ is on the order of $\hbar s$), and 
it has to be strongly interacting (because in a weakly interacting 
system the mean free path is large). It is also known that many 
of the model field theories that attain the bound in the strong 
coupling limit are scale invariant. All of these properties point 
to the unitary Fermi gas as a plausible candidate for a ``perfect
fluid''. 

 Almost ideal hydrodynamic flow in the unitary Fermi gas was first 
observed in \cite{OHara:2002}. Since then, a number of experiments 
have been performed that provide constraints on the shear viscosity
of the unitary gas \cite{Kinast:2004,Kinast:2004b,Kinast:2005,Altmeyer:2006,Altmeyer:2006b,Wright:2007,Clancy:2007,Riedl:2008}.
In this work we will provide an overview of the hydrodynamic analysis
of these experiments, and compare some of the estimates that have
been obtained. We emphasize the uncertainties of these results, and
point to improvements that need to be implemented.

\section{Scaling Flows}
\label{sec_scal_flow}

 We begin by studying the ideal (Eulerian) fluid dynamics of 
a non-relativistic gas in the normal phase. We will introduce 
dissipative effects in Sects.~\ref{sec_diss_de}-\ref{sec_diss_ns}.
In this contribution we will not discuss superfluid hydrodynamics. 
We will briefly comment on dissipative effects in the superfluid 
phase in Sect.~\ref{sec_diss_de}. The equations of continuity and of 
momentum conservation are given by 
\bea
\frac{\partial n}{\partial t} + \vec{\nabla}\cdot\left(n\vec{v}\right) 
 &=& 0 , \\
mn \frac{\partial \vec{v}}{\partial t} 
 + mn \left(\vec{v}\cdot\vec{\nabla} \right)\vec{v} &=& 
 -\vec{\nabla}P-n\vec{\nabla}V,
\eea 
where $n$ is the number density, $m$ is the mass of the atoms, $\vec{v}$ 
is the fluid velocity, $P$ is the pressure and $V$ is the external
potential. In the unitarity limit the equation of state at zero 
temperature is of the form
\be 
\label{P_uni}
P(n,T) = \frac{n^{5/3}}{m}f\left(\frac{mT}{n^{2/3}}\right)\, , 
\ee
where $f(y)$ is a universal function. We note that $y={\it const}\cdot
(T/T_F^{\it hom})$, where $T_F^{\it hom}=(3\pi^2n)^{2/3}/(2m)$ is the 
Fermi temperature of a homogeneous Fermi gas. In the high temperature 
limit, $y\gg 1$, we have $f(y)\simeq y$ and in the low temperature 
limit $f(y)\simeq (3\pi^2)^{2/3}\xi/5$, where the parameter $\xi=0.40(2)$ 
has been determined in quantum Monte Carlo calculations \cite{Carlson:2005}. 
Monte Carlo methods have also been used to determine $f(y)$ for all values 
of $y$ \cite{Bulgac:2008zz,Chevy:2009}. The critical temperature for 
superfluidity is $T_c/T_F^{\it hom}\simeq 0.15$, corresponding to 
$y_c\simeq 0.72$. An alternative representation of the pressure is 
\be 
P(\mu,T) = \mu^{5/2}m^{3/2} g\left(\frac{T}{\mu}\right)\, , 
\ee
where $g(z)$ is a universal function, related to $f(y)$ by thermodynamic
identities. In the high temperature limit $g(z)\simeq 2z^{5/2}e^{1/z}/
(2\pi)^{3/2}$ and in the low temperature limit $g(z)\simeq  2^{5/2}/
(15\pi^2\xi^{3/2})$. The density is 
\be 
\label{n_uni}
n(\mu,T) = \mu^{3/2}m^{3/2}h\left(\frac{T}{\mu}\right)\, , 
\hspace{0.5cm}
h(z)= \frac{5}{2}g(z)-zg'(z)\, . 
\ee
The high and low temperature limits of the function $h(z)$ are 
$h(z)\simeq 2z^{3/2}e^{1/z}/(2\pi)^{3/2}$ ($z\gg 1$) and $h(z)\simeq 
2^{3/2}/(3\pi^2\xi^{3/2})$ ($z\ll 1$).
The equilibrium distribution $n_0$ of a trapped atomic gas follows
from the hydrostatic equation $\vec{\nabla}P_0=-n_0\vec{\nabla}V$.
The trapping potential is approximately harmonic
\be 
V(x) = \frac{m}{2}\sum_i \omega_i^2 x_i^2 .
\ee
Using the Gibbs-Duhem relation $dP=nd\mu+sdT$ together with the 
fact that the equilibrium configuration is isothermal we can write 
the equation of hydrostatic equilibrium as $\vec{\nabla}\mu=-\vec{\nabla}
V$. This implies that the equilibrium density is $n_0(x)=n(\mu(x),T)$ with
\be
\label{LDA}
\mu(x)=\mu_0-V(x) = \mu_0 \left( 1- \sum_i \frac{x_i^2}{R_i^2}\right)\, ,
\hspace{0.5cm}
 R_i^2 = \frac{2\mu_0}{m\omega_i^2} \, . 
\ee
A scaling flow is a solution of the hydrodynamic equations in 
which the shape of the density distribution is preserved. Consider
the ansatz $n(x,t)=n(\mu(x,t),T(t))$ where 
\be 
\label{scal_flow_1}
\mu(x,t) = \mu_0(t)\left(
  1 - \frac{x^2}{R_x(t)^2} - \frac{y^2}{R_y(t)^2}
    - \frac{z^2}{R_z(t)^2} -  \frac{xy}{R_{xy}(t)} \right),
\ee
and  $T(t)/T(0)=\mu_0(t)/\mu_0(0)$. Without loss of generality
we have restricted the ansatz to rotations in the $xy$-plane. 
We note that the fluid remains isothermal during the expansion. 
Scale invariance implies that  properties of the fluid only 
depend on the dimensionless ratio $T/\mu$. For any given fluid 
element this ratio does not change during the expansion. In 
particular, if the fluid element was in the superfluid or 
normal phase initially, it will stay in that phase throughout 
the expansion. 

 The velocity field created by the scaling expansion in 
equ.~(\ref{scal_flow_1}) is linear in the coordinates.
We can write 
\be 
\label{scal_flow_2}
 \vec{v}(x,t) = \frac{1}{2}\vec{\nabla}
 \left( \alpha_x(t)x^2 + \alpha_y(t)y^2 + \alpha_z(t)z^2 
       +2\alpha(t)xy \right) + \Omega(t) \hat{z}\times \vec{x}.
\ee
The parameters $\alpha_i,\alpha$ and $\Omega$ are related to 
the parameters $R_i,R_{xy}$ and $\mu_0$ by the continuity equation. 
Remarkably, the continuity equation is independent of the universal 
function $h(z)$ in equ.~(\ref{n_uni}). Introducing the dimensionless 
scale parameters
\be 
\label{scal_par}
 \bar{\mu}(t) = \frac{\mu_0(t)}{\mu_0(0)},\hspace{0.5cm}
 b_i(t)=\frac{R_i(t)}{R_i(0)}, \hspace{0.5cm}
 a(t)  = \frac{R_x(0)^2}{R_{xy}(t)}\, ,
\ee
the continuity equation can be written as 
\bea
\label{cont_1}
\dot{\bar{\mu}} + \frac{2}{3} \bar{\mu} 
  \left( \alpha_x + \alpha_y + \alpha_z \right ) &=& 0\, ,  \\[0.1cm]
\dot a + \frac{2(\alpha-\Omega)}{b_x^2}
       + \frac{2(\alpha+\Omega)}{\lambda^2b_y^2}
       + a (\alpha_x+\alpha_y) &=& 0 \, ,  \\
\dot b_x - b_x\alpha_x 
       - \frac{b_x^3 a}{2} (\alpha+\Omega ) &=& 0 \, ,  \\
\dot b_y - b_y\alpha_y 
       - \frac{b_y^3 \lambda^2 a}{2} (\alpha-\Omega ) &=& 0 \, , \\
\label{cont_5}
\dot b_z - b_z\alpha_z  &=& 0 \, , 
\eea
where $\lambda=R_y(0)/R_x(0)=\omega_x/\omega_y$. These equations can 
be solved directly in the case that there is no rotation, $a(t)=0$. 
Then $\alpha=\Omega=0$ and
\be
 (\alpha_x,\alpha_y,\alpha_z) = 
           \left(\frac{\dot b_x}{b_x},\frac{\dot b_y}{b_y},
                 \frac{\dot b_z}{b_z}\right), 
\hspace{0.5cm} 
\bar{\mu} = \frac{1}{(b_x b_y b_z)^{2/3}}\, . 
\ee
The velocity field is a simple ``Hubble flow'', $\vec{v}= 
(\alpha_x x,\alpha_y y,\alpha_z z)$. Finally, we note that the
entropy density is given by $s=(m\mu)^{3/2}g'(T/\mu)$. Since 
the entropy density has the same functional form as the particle 
density we conclude that, in the case of scaling flows, the 
continuity equation implies entropy conservation, 
\be
\label{s_cons}
\frac{\partial s}{\partial t}+\vec{\nabla}\cdot(\vec{v}s)=0\, . 
\ee

\section{Elliptic Flow}
\label{sec_ell}

 The simplest scaling flow is the expansion of the cloud after
the trapping potential is removed \cite{Menotti:2002}. Since 
the cloud remains isothermal the Euler equation can be derived
using the Gibbs-Duhem relation $dP=nd\mu$. This implies that the 
equation of motion is independent of the universal function $f(y)$ 
defined in equ.~(\ref{P_uni}). We get  
\be 
\label{euler_flow}
 \ddot b_i = \frac{\omega_i^2}{(b_xb_yb_z)^{2/3}}\frac{1}{b_i}\, ,  
\ee
The total energy of the expanding system is given by the 
sum of internal energy and kinetic energy,  
\be 
\label{e_tot}
 E= E_{int} + E_{kin} 
  = \int d^3x\,\left( {\cal E}(x) + \frac{1}{2}\,mn\vec{v}^{\,2}\right) \, . 
\ee
For the Fermi gas at unitarity the energy density ${\cal E}$ is related 
to the pressure by ${\cal E}=\frac{3}{2}P$. We find
\be
\label{E_flow}
 E= E_{int}(0) \left\{ \frac{1}{(b_x b_y b_z)^{2/3}} 
  + \frac{1}{3}\left( \frac{\dot b_x^2}{\omega_x^2}
                     +\frac{\dot b_y^2}{\omega_y^2}
                     +\frac{\dot b_z^2}{\omega_z^2}\right)\right\},
\ee
where $E_{int}(0)$ is the internal energy at $t=0$. Conservation of energy 
immediately follows from the equation of motion, equ.~(\ref{euler_flow}).
We note that the equation of hydrostatic equilibrium, $\vec{\nabla}P =
-n\vec{\nabla}V$, implies the Virial theorem $\langle {\cal E}\rangle 
=\langle V\rangle$ \cite{Thomas:2005}, where $\langle V \rangle$ denotes 
the integral of the potential energy over the trap. This means that the 
total energy of the trapped gas is $E_0=2E_{int}(0)$, where the factor 2 
is due to the contribution of the potential energy. 

 We are interested in an axially symmetric trap with $\omega_y=\omega_z
=\omega_\perp$ and $\omega_x=\lambda\omega_\perp$. In this case we end 
up with two coupled equations for $b_\perp$ and $b_x$. If $\lambda\gg 1$ 
the evolution in the transverse direction is much faster and the equation 
for $b_\perp$ can be approximately decoupled, 
\be 
 \ddot b_\perp = \frac{\omega_\perp^2}{b_\perp^{7/3}}\, .
\ee
This equation has to be integrated numerically. The behavior 
at early and late  times can be found analytically. We get
\be 
\label{b_perp_analyt}
 b_\perp(t) \simeq\left\{ \begin{array}{cl}
 1+\frac{1}{2}\,\omega_\perp^2t^2 + O(t^4)\;\; 
                           & \omega_\perp t\ll 1 \, , \\
 \frac{\omega_\perp t}{\sqrt{\gamma}} + c_0 + O(t^{-1/3})  
                           & \omega_\perp t\gg 1 \, , 
\end{array}\right. \, ,
\ee
where $\gamma=2/3$ and $c_0$ is a constant that can be determined
by matching the early and late time behavior. Numerically, we find
$c_0\simeq -1.3$. For the longitudinal expansion the early time 
behavior is $b_x(t)\simeq 1+(\lambda\omega_\perp t)^2/2$, and at 
late times $b_x(t)\simeq {\it const}\cdot \lambda^2\omega_\perp t$. 

 The signature effect of hydrodynamics is that transverse
pressure gradients cause the transverse radius to expand
much faster than the longitudinal radius. This means that
the two radii will eventually cross. This happens at a time
\be 
\label{t_cross}
 t_{\it cross} = \frac{\sqrt{\gamma}}{\omega_x} 
  \left( 1+ O(\lambda) \right)\, .
\ee
We note that the crossing time only depends on the trap parameters, 
and is independent of the initial energy or the number of particles. 
We also note that at $t\simeq t_{\it cross}$ the expansion is still 
two-dimensional, that means the volume of the system grows as ${\it vol}
\sim t^2$. The expansion becomes three-dimensional, ${\it vol}\sim t^3$,
at $t_{\it 3d}\sim (\lambda^2\omega_\perp)^{-1}$. 

\subsection{Energy dissipation}
\label{sec_diss_de}

 We wish to understand how the expansion is affected by dissipation.
The energy momentum tensor of a dissipative fluid is $\Pi_{ij}=P\delta_{ij} 
+mnv_iv_j+\delta\Pi_{ij}$ with 
\be 
\label{del_pi_ns}
 \delta \Pi_{ij} = \eta \left(\nabla_i v_j+\nabla_j v_i
  - \frac{2}{3}\delta_{ij}\nabla\cdot v\right) 
  + \zeta \delta_{ij} \left( \nabla\cdot v\right) \, . 
\ee
The energy current is $j_i^\epsilon = v_i(w+\frac{1}{2}mnv^2)+\delta
j_i^\epsilon$ with $w={\cal E}+P$ and $\delta j_i^\epsilon = \delta
\Pi_{ij} v_j - \kappa \nabla_i T$. The unitary gas is scale invariant 
and $\zeta=0$ \cite{Son:2005tj}. Also, for an isentropic scaling expansion 
the temperature remains independent of position, and there is no contribution 
from the thermal conductivity $\kappa$. We will therefore concentrate on 
the role of shear viscosity. 

 Since the shear viscosity is small, we can take it into account 
perturbatively. The simplest idea it compute the amount of kinetic 
energy that is converted to heat. We have 
\be 
\dot{E} = -\frac{1}{2}\int d^3x\,\eta 
\left( \nabla_i v_j+\nabla_j v_i
  - \frac{2}{3}\delta_{ij}\nabla\cdot v\right)^2 . 
\ee 
For the scaling expansion given in equ.~(\ref{scal_flow_2}) 
the result is particularly simple. We get
\be 
\label{e_dot}
\dot{E} = -\frac{4}{3} \left(\frac{\dot b_\perp}{b_\perp}
 -\frac{\dot b_x}{b_x}\right)^2 \int d^3x\, \eta(x). 
\ee
The total energy dissipated is given by the integral of 
equ.~(\ref{e_dot}) over time. We first show that the spatial 
integral over $\eta(x)$ does not depend on time. In the local 
density approximation $\eta(x)=\eta(\mu(x),T)$. Scale invariance 
implies that 
\be 
\label{eta_uni}
 \eta(\mu,T) = n(\mu,T) \alpha_n\left(\frac{T}{\mu}\right)\, ,
\ee
where $\alpha_n(z)$ is a universal function, and we have set 
$\hbar=1$. In order to compare with the string theory bound 
it is also useful to define $\eta(\mu,T) = s(\mu,T)\alpha_s(T/\mu)$,
where we have also set $k_B=1$. We can write
\be 
 \int d^3x\, \eta(x) = N\langle \alpha_n\rangle \, ,
\ee
where 
\be
 \langle \alpha_n\rangle 
 = \frac{1}{N}\int d^3x \,  n(x,t)\,
     \alpha_n \left( \frac{T(t)}{\mu(x,t)}\right)
 = \frac{1}{N}\int d^3x \,  n_0(x)\,
     \alpha_n \left( \frac{T_0}{\mu(x,0)}\right)\, 
\ee
is an average of $\alpha_n$ over the initial density distribution. 
Analogously, we can write the integral over $\eta(x)$ as $S\langle
\alpha_s\rangle$, where $S$ is the total entropy and $\langle\alpha_s
\rangle$ is an average of $\alpha_s$ over the initial entropy density. 

 The time integral over $(\dot b_\perp/b_\perp-\dot b_x/b_x)^2$ 
is dominated by the regime $\omega_\perp t\sim 1$ and converges 
rapidly -- the integral reaches 80\% of its asymptotic value at
$t_{\it diss}\simeq 5.9\,\omega_\perp^{-1}$. In the limit $\lambda\ll 1$ 
we can neglect the contribution from $\dot b_x$. On dimensional 
grounds the integral over $(\dot b_\perp/b_\perp)^2$ must be 
proportional to $\omega_\perp$. The constant of proportionality 
can be determined numerically. We find
\be
\label{del_e_int}
 \int_0^\infty dt\, \left(\frac{\dot b_\perp}{b_\perp}\right)^2 
 = 0.87 \,\omega_\perp \, . 
\ee
We can now compute the ratio $\Delta E/E_{int}$ of the dissipated 
energy to the initial internal energy of the system. In order to 
express the result in terms of experimentally measured quantities
it is useful to introduce the energy $E_F=N\epsilon_F$ where 
$\epsilon_F=\bar{\omega}(3N)^{1/3}$ is the Fermi energy of the 
trapped gas and $\bar{\omega}=(\omega_x\omega_y\omega_z)^{1/3}$. 
We find
\be 
\label{delE_lin_beta}
\frac{\Delta E}{E_{int}(0)} = 
 -\frac{8}{3}\cdot 0.87 \cdot \beta
 = -2.32 \cdot \beta
\ee
where the parameter $\beta$ is defined given by
\be 
\label{beta}
 \beta = 
 \frac{\langle\alpha_n\rangle}{(3N\lambda)^{1/3}}
 \frac{1}{(E_0/E_F)}  = 
 \frac{\langle\alpha_s\rangle}{(3N\lambda)^{1/3}}
 \frac{(S/N)}{(E_0/E_F)}\, . 
\ee
Dissipation slows down the transverse expansion of the system. 
For $(\omega_\perp t)\gg 1$ we have $(\delta\dot{b}_\perp/\dot{b}_\perp)
=(\Delta E/E)/2$ and, up to terms that are higher order in 
$\lambda$, the change in the crossing time is directly related 
to the change in the expansion rate, $(\delta t/t)_{\it cross}=
(\delta\dot{b}_\perp/\dot{b}_\perp)$.

\begin{figure}[t]
\bc\includegraphics[width=10cm]{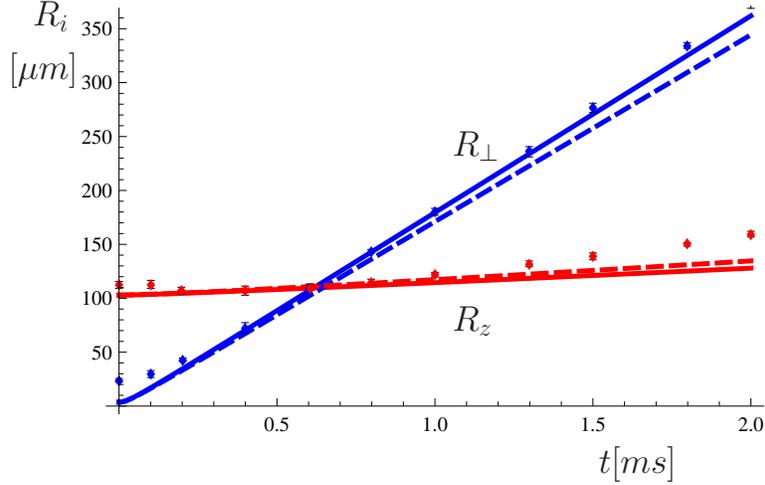}\ec
\caption{\label{fig_flow}
Expansion of the transverse and longitudinal radii
after release from a harmonic trap. The data points are taken
from \cite{OHara:2002}. The solid and dashed lines correspond
to solutions of the Navier-Stokes equation with $\langle \alpha_s
\rangle =0$ (solid lines) and $\langle \alpha_s\rangle =0.5$ 
(dashed lines).}
\end{figure}

 The thermodynamic quantities $S/N$ and $E_0/E_F$ as a function 
of $T/T_F$ were determined experimentally in \cite{Luo:2008}. 
Just above the critical temperature $S/N\simeq 2.2$ and $E_0/E_F
\simeq 0.83$. The double ratio $[(S/N)/(E_0/E_F)]$ is only 
weakly dependent on $T$, changing by less than 15\% between 
$T_c$ and $4T_c$. In the flow experiment carried out by O'Hara
et al.~\cite{OHara:2002} the cloud contained $N=2\cdot 10^5$ 
atoms and the asymmetry parameter was $\lambda=0.045$. The 
predicted sensitivity of the crossing time to dissipative 
effects is 
\be 
\label{del_t_cr}
 \left(\frac{\delta t}{t}\right)_{\it cross} 
 = 0.008
  \left( \frac{\langle \alpha_s\rangle}{1/(4\pi)}\right)\,
  \left( \frac{2\cdot 10^5}{N} \right)^{1/3}
  \left( \frac{0.045}{\lambda} \right)^{1/3}
  \left( \frac{S/N}{2.2} \right)\,
  \left( \frac{0.83}{E_0/E_F} \right)\, . 
\ee
For $\langle\alpha_s\rangle = 1/(4\pi)$ this is at the limit of what 
can be resolved experimentally, but for $\langle\alpha_s\rangle=0.5$ 
the effect reaches about 5\%. An example is shown in Fig.~\ref{fig_flow}. 
The solid lines show the solution of the Euler equation (\ref{euler_flow}), 
and the dashed lines show a solution of the Navier-Stokes equation (see 
Sect.~\ref{sec_diss_mom}) with $\langle\alpha_s\rangle=0.5$. The main 
effect of shear viscosity is a suppression of the transverse expansion 
of the system. We find $(\delta t/t)_{\it cross}=6.5\%$, in fairly good
agreement with the estimate $(\delta t/t)_{\it cross}=5\%$ from 
equ.~(\ref{del_t_cr}). 

 The best fit to the data is provided by ideal hydrodynamics with 
$\langle \alpha_s\rangle=0$. This is probably related to the fact that 
the data were taken significantly below $T_c$, at $T/T_F=0.13\pm 0.05$. 
In this regime the system is described by two-fluid hydrodynamics. The 
superfluid component has no shear viscosity but the viscosity of the 
normal component becomes very large as $T/T_F\to 0$ \cite{Rupak:2007vp}.
In a finite system, however, the large viscosity of the normal phase is 
likely to be suppressed by relaxation time effects, see 
Sect.~\ref{sec_diss_relax}. As a consequence one observes perfect 
superfluid hydrodynamics. The data in Fig.~\ref{fig_flow} show some 
deviations from hydrodynamics at very early and very late times. 
Discrepancies at early times are probably related to experimental
resolution \cite{OHara:2002}, while the differences at late times
may be connected to the breakdown of hydrodynamics in the late
stages of the expansion.

 We can also compute the amount of entropy generated by dissipative 
effects. Using $dS=dQ/T$ we find
\be 
 \frac{\Delta S}{N} =  \frac{4}{3}
  \frac{\langle\alpha_n\rangle}{(3N\lambda)^{1/3}}
   \frac{1}{(T_0/T_F)} \, I_{S}\, 
\ee
with
\be 
 I_{S} = \omega_\perp^{-1} \int^{\tau}_0 dt\, 
  b_\perp^{-2/3} \left(\dot b_\perp\right)^2 . 
\ee
For $\tau\simeq t_{\it diss}$ we find $I_s\simeq 2.6$ and the produced
entropy is small, $(\Delta S/N)\simeq 0.27$ for the conditions given
above. However, the integral diverges as $I_s\sim (\omega_\perp \tau)^{1/3}$
for $\tau\to\infty$. This result is not reliable since we expect 
hydrodynamics to break down at late times, see Sect.~\ref{sec_break}.

\subsection{Moments of the Navier-Stokes equation}
\label{sec_diss_mom}

 It is clearly desirable to study the role of dissipation more directly 
by solving the Navier-Stokes equation. The Navier-Stokes equation differs 
from the Euler equation by an extra term on the right hand side, 
\be 
\label{ns}
 mn \left(\frac{\partial v_i}{\partial t} 
 +  \left(\vec{v}\cdot\vec{\nabla}\right)v_i \right) = 
  -\nabla_i P  - \nabla_j\, \delta \Pi_{ij} . 
\ee
We will assume that the viscosity is small, so that derivatives with 
respect to thermodynamic variables can be computed at constant entropy.
We will also assume that the entropy conservation equation, 
equ.~(\ref{s_cons}), is not modified. Physically, this implies that 
we assume that there is a reservoir that removes the heat generated
by dissipative effects. In this case, the only correction to the 
equations of hydrodynamics is the viscous force in the Navier-Stokes
equation.

 In general the inclusion of the Navier-Stokes term will break
the simple scaling form of the flow. The Navier-Stokes equation
also depends on the functional form of the pressure and the 
viscosity, that means we have to specify the functions $f(y)$ 
in equ.~(\ref{P_uni}) and $\alpha_n(z)$ in equ.~(\ref{eta_uni}). 
A simple approach that avoids extensive numerical work as well
as model assumptions about $f(y)$ and $\alpha_n(z)$ is to take 
moments of the Navier-Stokes equation. Consider the linear 
moments
\be 
\label{ns_mom}
 m\int d^3x\, x_k n(x) \left(
  \frac{\partial v_i}{\partial t} 
 +  \left(\vec{v}\cdot\vec{\nabla}\right)v_i\right)  
 = -\int d^3x\, x_k\Big( \nabla_i P 
        + \nabla_j\, \delta \Pi_{ij}\Big) \, , 
\ee
with $k=1,2,3$. Since the velocity field is linear in 
the coordinates we find that the ideal fluid terms involve 
second moments of the density. These moments are related to 
the potential energy in a harmonic trap and, by the virial 
theorem, to the total energy of the system. The Navier-Stokes 
term can be integrated by parts and is proportional
to the integral over $\eta(x)$. As a consequence, the 
first moment of the Navier-Stokes equation depends only 
on the parameter $\beta$ defined in equ.~(\ref{beta}).
We get
\bea
\label{ns_mom_1} 
\ddot b_\perp  &=& \frac{\omega_\perp^2}
   {(b_\perp^2 b_x)^{2/3} b_\perp}
   -  \frac{2\beta\omega_\perp}{b_\perp}
      \left( \frac{\dot b_\perp}{b_\perp} 
                - \frac{\dot b_x}{b_x} \right) \\
\label{ns_mom_2}
\ddot b_x  &=& \frac{\omega_x^2}
   {(b_\perp^2 b_x)^{2/3} b_x}
   +  \frac{4\beta\lambda\omega_x }{b_x}
      \left( \frac{\dot b_\perp}{b_\perp} 
                - \frac{\dot b_x}{b_x} \right) .
\eea
\begin{figure}[t]
\bc\includegraphics[width=9cm]{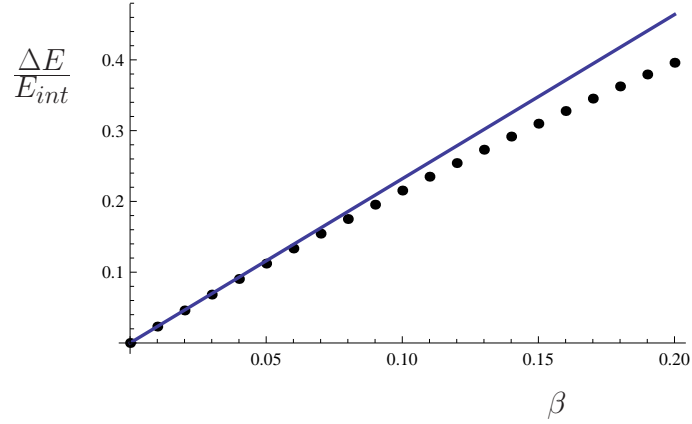}\ec
\caption{\label{fig_dele}
This figure show the ratio $(\Delta E)/E_{int}$ of the dissipated 
energy to the initial internal energy as a function of the parameter
$\beta$ defined in equ.~(\ref{beta}). The dots show the result of
a numerical solution of the Navier-Stokes equation (\ref{ns_mom_1},
\ref{ns_mom_2}) in the limit $\omega_z/\omega_\perp\to 0$, and the 
line shows the estimate given in equ.~(\ref{delE_lin_beta}).}
\end{figure}
These equations of motion are consistent with the result in the 
previous section. We can compute the amount of energy dissipated 
from equ.~(\ref{E_flow}) and (\ref{ns_mom_1},\ref{ns_mom_2}). 
We find
\be 
\label{e_dot_2}
 \dot E = -\frac{8}{3}\beta E_{int}(0) 
  \left( \frac{\dot b_\perp}{b_\perp}-\frac{\dot b_x}{b_x}\right)^2\, . 
\ee
We note that $b_\perp(t)$ and $b_z(t)$ are solutions of the Navier-Stokes 
equation and have an implicit dependence on $\beta$. As long as this 
dependence is smooth, $b_i(t,\beta)\to b_i(t,0)$ as $\beta\to 0$, 
equ.~(\ref{e_dot_2}) reduces to equ.~(\ref{e_dot}) at leading order 
in $\beta$. Since typical values of $\beta$ are quite small, we 
expect the estimates in the previous section to be very
accurate. This is studied in more detail in Fig.~\ref{fig_dele}.
We observe that the dissipated energy $(\Delta E)/E$ is very linear 
in $\beta$ even for values of $(\Delta E)/E$ as large as 25\%. 
We note that because of turbulence solutions of the Navier-Stokes
equation do not in general approach solutions of the Euler equation
in the limit that the shear viscosity goes to zero. Turbulence is 
not present in our analysis because we do not consider small 
fluctuations. We also note that there is no continuous forcing 
in the case of an expanding gas and it is not clear whether 
turbulence can develop even if fluctuations are included. We 
will estimate the Reynolds number of the flow in Sect.~\ref{sec_break}.

\subsection{Scaling solution of the Navier-Stokes equation}
\label{sec_diss_ns}

 In this section we discuss a specific model for the density 
dependence of the shear viscosity that preserves the scaling 
nature of the flow even if the viscosity is not zero. This 
model allows to compute the local amount of heat that is 
generated by dissipation, and to understand some of the 
shortcomings of the method discussed in Sections \ref{sec_diss_de} 
and \ref{sec_diss_mom}. Consider 
\be 
\label{eta_scal}
 \eta(n,T) = \eta_0 (mT)^{3/2} 
   + \eta_1 \frac{P(n,T)}{T} ,
\ee
where $\eta_{0,1}$ are constants and $P(n,T)$ is the pressure.
The first term dominates in the low density, high temperature
limit. This is the regime in which a kinetic description in terms
of weakly coupled atoms is applicable. Kinetic theory gives
\cite{Bruun:2005,Bruun:2006}
\be 
\label{eta_kin}
\eta_0 = \frac{15}{32\sqrt{\pi}}\, . 
\ee
The second term dominates in the high density, low temperature
regime. The functional form of this term is not motivated 
by kinetic theory. We note, however, that $\eta/n$ has a 
minimum as a function of $T$, as expected on theoretical 
\cite{Rupak:2007vp} and phenomenological 
grounds \cite{Schafer:2009dj}. 

\begin{figure}[t]
\bc\includegraphics[width=9cm]{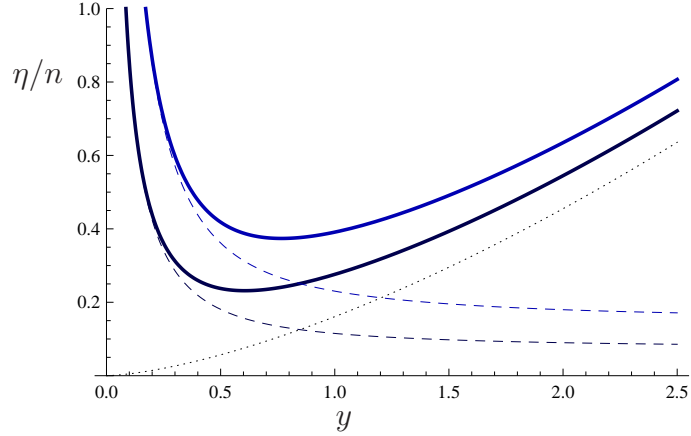}\ec
\caption{\label{fig_alpha_n}
Ratio $\eta/n$ as a function of $y=(mT)/n^{2/3}$ for the model
defined in equ.~(\ref{eta_scal}). The two curves correspond to 
(from bottom to top) $\eta_1=1/(4\pi),2/(4\pi)$ with $\eta_0=
15/(32\sqrt{\pi})$. The dashed line shows the contribution from 
$\eta_1$, which is the term that contributes directly to the 
Navier-Stokes equation, and the dotted line is the contribution
from $\eta_0$. Note that the critical point for the onset of 
superfluidity is $y_c\simeq 0.72$.}
\end{figure}

 The model given in equ.~(\ref{eta_scal}) has two remarkable 
features: First, the $\eta_0$ term does not contribute to 
the Navier-Stokes equation at all. The Navier-Stokes term
$\nabla_j[\eta_0(mT)^{3/2}(\nabla_i v_j+ \ldots)]$ vanishes 
since both $T$ and $\nabla_iv_j$ are constant. Second, the $\eta_1$ 
term preserves the scaling flow. Using $T,\nabla_iv_j\sim {\it 
const}$ we see that $\nabla_j[\eta_1 P(n,T)/T$ $(\nabla_i v_j+\ldots)]$ 
scales like the contribution from the pressure of an ideal fluid, 
$\nabla_i P(n,T)$. We get
\bea
\label{ns_scal_1} 
\ddot b_\perp  &=& \frac{\omega_\perp^2}
   {(b_\perp^2 b_x)^{2/3} b_\perp}
   -  \frac{2\eta_1\omega_\perp^2}{3T_0b_\perp}
      \left( \frac{\dot b_\perp}{b_\perp} 
                - \frac{\dot b_x}{b_x} \right) \\
\label{ns_scal_2}
\ddot b_x  &=& \frac{\omega_x^2}
   {(b_\perp^2 b_x)^{2/3} b_x}
   +  \frac{4\eta_1\omega_x^2}{3T_0b_x}
      \left( \frac{\dot b_\perp}{b_\perp} 
                - \frac{\dot b_x}{b_x} \right) .
\eea
We observe that these equations are identical to the moment 
equations (\ref{ns_mom_1},\ref{ns_mom_2}) with $\beta = \eta_1
\omega_\perp/(3T_0)$. This is not a surprise -- the $\eta_1$ contribution
to $\eta(n,T)$ vanishes as $n\to 0$ and the assumptions underlying 
the moment method are satisfied. The $\eta_0$ term, on the other hand, 
does not vanish as $n\to 0$, and it cannot be included in the 
moment equations (it makes an infinite contribution to the 
integral over $\eta(x)$). 

 Using the identification $\beta = \eta_1 \omega_\perp/(3T_0)$ 
we can write
\be 
\label{beta_eta_1}
\beta = \frac{\eta_1}{3(3\lambda N)^{1/3}}
        \frac{1}{(T_0/T_F)}\, ,  
\ee
which shows that any bound on $\langle\alpha_n\rangle$ obtained
using the methods of Sect.~\ref{sec_diss_mom} can be translated 
into an estimate of $\eta_1$, $\eta_1=3(T_0/E_0)\langle\alpha_n\rangle$. 
Near $T_c$ this implies that $\eta_1\simeq 0.76\langle\alpha_n\rangle$.
We note that the relation between $\eta_1$ and $\langle\alpha_n\rangle$
is precisely what one obtains if the trap average of $\eta(x)$ is
computed from the $\eta_1$-term only. The situation is more 
complicated if the contribution from $\eta_0$ is taken into account. 
The ratio $\eta/n$ is given by 
\be 
 \frac{\eta(n,T)}{n} = \eta_0 y^{3/2} + \frac{\eta_1}{y} f(y)\, 
\ee
with $y=(mT)/n^{2/3}$. Since $f(0)={\it const}$ and $f(y)\simeq y$ 
for $y\gg 1$ this function has a minimum, see Fig.~\ref{fig_alpha_n}.
The figure also shows that $(\eta/n)_{\it min}$ receives significant 
contributions from $\eta_0$. It is clearly unsatisfactory that 
our analysis has no sensitivity to this term. We will return 
to this issue in Sect.~\ref{sec_diss_relax}. 

 Using the explicit form of $\eta(n,T)$ we can also address the 
question where the energy is being dissipated and how much 
reheating is taking place. We first consider the contribution from 
$\eta_1$. The energy dissipated is 
\be 
\label{dot_eps_eta_1}
 \dot{\cal E} = -\frac{4\eta_1}{3}
   \left(\frac{\dot{b}_\perp}{b_\perp}\right)^2
   \frac{P(n,T)}{T}\, .
\ee
For a Fermi gas at unitarity the energy density is related to the 
pressure by ${\cal E}(n,T)=(3/2)P(n,T)$. Equ.~(\ref{dot_eps_eta_1}) 
implies that the energy dissipated is proportional to the local
internal energy density. The source of the dissipated energy is 
the reduction in the kinetic energy density relative to its value 
in ideal hydrodynamics. The local kinetic energy density is 
\be
{\cal E}_{\it kin}= \frac{m}{2}n 
  \left(\frac{\dot{b}_\perp}{b_\perp}\right)^2 x_\perp^{\, 2}\, . 
\ee 
Since the kinetic energy density differs from the spatial distribution 
of the dissipated energy there has to be a dissipative contribution to 
the energy current. This current is given by $\delta \vec{j}^{\;
\epsilon}= (0,\delta j^{\,\epsilon}_y,\delta j^{\,\epsilon}_z)$ with 
\be
\delta j^{\,\epsilon}_z=v_z\delta\Pi_{zz}
 = -z \frac{2\eta_1P(n,T)}{3T}  
   \left(\frac{\dot{b}_\perp}{b_\perp}\right)^2 \, , 
\ee
and $\delta j^{\,\epsilon}_y=\delta j^{\,\epsilon}_z
(z\leftrightarrow y)$. The dissipative current flows from the 
outer edge of the cloud, where the kinetic energy is peaked, to
the center of the cloud, where the pressure is largest. 

 Energy dissipation leads to reheating. The change in temperature 
is $\Delta T =(\Delta {\cal E})/c_V$. The time evolution of the 
temperature is governed by 
\be 
\dot T = -\frac{4T_0}{3\,b_\perp^{4/3}}
    \left(\frac{\dot{b}_\perp}{b_\perp}\right)
  + \frac{\eta_1 P}{c_VT} 
    \left(\frac{\dot{b}_\perp}{b_\perp}\right)^2 \, , 
\ee
where the first term is related to the adiabatic expansion
of the system, and the second term is the dissipative correction. 
Note that if $c_V\sim {\cal E}/T$, which is the case in the high 
temperature limit, then reheating will preserve the fact that the 
cloud is isothermal. In general the behavior of the specific heat 
is more complicated and dissipation produces a temperature gradient. 
The relative importance of reheating is governed by the parameter 
$(\eta_1 \omega_\perp/T_0)(P/(c_VT))$. In the high temperature limit 
we can use $P\sim c_VT$ and this expression reduces to the parameter 
$\beta$ defined in equ.~(\ref{beta_eta_1}). Reheating becomes 
important at a time $\omega_\perp t \sim \beta^{-3}$. Since $\beta$ 
is typically very small, this occurs very late during the evolution 
of the system. 

 A similar analysis of the effects of $\eta_0$ leads to a number 
of puzzles. The energy dissipated is independent of density, and 
the total energy dissipated over all space is infinite. There is 
no change in the kinetic energy, and the source of the dissipated
energy is the viscous correction to the energy current. This 
current flows into the system from spatial infinity. The relative 
importance of reheating is governed by the parameter 
$(\eta_0\omega_\perp/T_0)((mT)^{3/2}/n)$, which is always 
large in the dilute region of the cloud. 

\subsection{Breakdown of hydrodynamics}
\label{sec_break}

 The constant term $\eta\sim \eta_0 (mT)^{3/2}$ in the shear 
viscosity dominates in the dilute outer regions of the cloud, and
the difficulty in understanding the effects of this term must be 
related to the breakdown of hydrodynamics in the dilute regime. 
A standard criterion for the applicability of hydrodynamics is the
condition that the Knudsen number ${\it Kn}=l_{\it mfp}/L$, the ratio 
of the mean free path to the system size, is much less than one. In
the dilute regime the mean free path is given by 
\be 
 l_{\it mfp} = \frac{1}{n\sigma} = 
  \frac{3}{4\pi}\frac{mT}{n}\, .
\ee
The density is given by equ.~(\ref{n_uni}). In the dilute regime 
we can use the high temperature limit of $h(z)$, but the scaling 
arguments in the following are independent of the functional form
of $h(z)$. For a comoving observer the density scales as $n\sim 
(m\mu)^{3/2}$, and the mean free path scales as $l_{\it mfp}\sim 
T/(m^{1/2}\mu^{3/2})$. The evolution of $T$ and $\mu$ is governed 
by the scaling relations discussed in Sect.~\ref{sec_scal_flow}.
We may use, in particular, that $T/\mu\sim{\it const}$ and $\mu
\sim \mu(0)/(b_\perp^2 b_x)^{2/3}$. We conclude that in a 
comoving fluid cell 
\be 
{\it Kn} = \frac{l_{\it mfp}}{L}
 \sim \left(\frac{b_x}{b_\perp}\right)^{1/3}\, . 
\ee
During the two-dimensional expansion the Knudsen number is 
dropping, which implies that the hydrodynamic description is 
becoming more accurate. In the late, three-dimensional stage, 
the Knudsen number is constant.

 A more accurate criterion can be obtained by using a characteristic
length or time scale derived from the flow profile. Hydrodynamics
is based on a derivative expansion of the energy momentum tensor, 
and the validity of hydrodynamics requires that $\delta\Pi_{ij}$ 
is small compared to the ideal fluid stress tensor. Consider 
the ratio of the moments of the ideal and dissipative terms
on the RHS of the Navier-Stokes equation 
\be 
\frac{\left\langle x_k \nabla_k P\right\rangle}
{\left\langle x_k \nabla_j \delta \Pi_{kj}\right\rangle}
 = \frac{\left\langle P\right\rangle}
        {\left\langle \frac{4}{3}\eta(\nabla_k v_k)\right\rangle}\, 
\ee
where $\langle.\rangle$ denotes an integral over $d^3x$ and the 
index $k$ is fixed. The ratio $(\eta/P)(\nabla\cdot v)$ has a 
simple interpretation in kinetic theory. For a dilute gas
$\eta\sim n p l_{\it mfp}\sim \rho u^2\tau_{\it mft}$ and 
$P\sim \rho u^2$, where $n$ is the particle density, $\rho$ is 
the mass density, $p$ is the average quasi-particle momentum, 
$u$ the average velocity, and $\tau_{\it mft}$ the mean free time. 
The ratio $\nabla\cdot v\sim \tau_{\it exp}^{-1}$ defines a 
characteristic expansion time. The quantity
\be
\label{fr_out}
\frac{\eta}{P}(\nabla\cdot v) 
  \sim \frac{\tau_{\it mft}}{\tau_{\it exp}}
\ee
measures the ratio of the mean free time over the expansion time. 
Hydrodynamics is valid if $\tau_{\it mft}\ll\tau_{\it exp}$. 
We observe that for $\eta\sim P$ the freezeout criterion is 
independent of position and only a function of time. We get
\be 
\label{fr_out_eta_1}
\frac{\eta}{P}(\nabla_z v_z)  
=  \frac{\eta_1}{T_0}
   \left(b_x b_\perp \right)^{1/3}\dot b_\perp
\simeq \frac{\eta_1}{(3N)^{1/3}\lambda^{1/3}}\frac{1}{(T_0/T_F)}
 \left(\omega_\perp t\right)^{1/3}\, ,   
\ee
where we have assumed that the expansion is two-dimensional. We note 
that the relevant parameter is the quantity $\beta$ defined in 
equ.~(\ref{beta_eta_1}). Freezeout occurs at $(\omega_\perp t_{\it fr})
\sim \beta^{-3}$. For typical values of $\beta$ we find that $t_{\it fr}
\gg t_{\it cross} \gg t_{\it diss}$, where $t_{\it cross}\sim 
(\omega_\perp\lambda)^{-1}$ is the crossing time, and $t_{\it diss}
\sim 5.9\,\omega_\perp^{-1}$ is the characteristic time for dissipative 
effects. 

 The freezeout time defined by equ.~(\ref{fr_out_eta_1}) is very 
long, and the physical freezeout is determined by the viscous
effects in the dilute part of the cloud. In the case of a 
spatially constant shear viscosity we find
\be 
\label{fr_out_eta_0}
\frac{\eta}{P}(\nabla_z v_z)  
=  \frac{\eta_0(mT)^{3/2}}{P}
   \left(\frac{\dot b_\perp}{b_\perp}\right)
\simeq  \frac{45\pi}{8\sqrt{2}}
  \frac{(T_0/T_F)^2}{(3\lambda N)^{1/3}}\,
   b_\perp^{1/3}\dot b_\perp  
  \exp\left(\sum_i\frac{x_i^2}{b_i^2 \bar{R}^2_i}
  \right)\, ,
\ee
where we have used $P=nT$ as well as the low density (high temperature) 
limit of $n_0(x)$, see equ.~(\ref{n_uni}). The radius parameter 
$\bar{R}_i$ is defined as $\bar{R}_i^2=2T_0/(m\omega_i^2)$. The 
condition $(\eta/P)(\nabla_z v_z)$ determines a freezeout surface 
$x_{\it fr}(t)$. This surface is initially at $x_i\gg R_i$, but it 
moves inward as time increases and reaches the origin at a time 
$t_{\it fr}\sim \omega_\perp^{-1} (3\lambda N)(T_F/T_0)^6$. This time 
is also parametrically very long, but the freezeout time at a 
characteristic distance $x_i\simeq b_i \bar{R}_i$ is significantly 
smaller.

 Finally, we wish to mention one more quantity that characterizes
a viscous flow. The Reynolds number ${\it Re}$ is defined as the 
ratio of inertial and viscous forces in the system. In the case 
of a scaling flow with $\eta\sim P$ this ratio is independent of 
position and only a function of time. We find 
\be 
{\it Re} = \frac{T_0}{\eta_1\omega_\perp^2}\, b_\perp\dot{b}_\perp
 \simeq \frac{\omega_\perp t}{\beta} \, . 
\ee
The Reynolds number is zero initially, but it grows quickly, reaching
${\it Re}\simeq \beta^{-1}$ at $(\omega_\perp t)=1$. For typical 
experimental parameters $\beta^{-1}\sim 100$, which is large but 
not large enough to cause instabilities. At later times even larger
values of ${\it Re}$ are reached, but at these late times the 
system is simply free streaming. A constant contribution to the 
viscosity does not lead to a viscous force, and does not directly 
contribute to the Reynolds number. 

\subsection{Relaxation time approach}
\label{sec_diss_relax}

 The discussion in the previous section does not fully resolve
the problems caused by the dilute regions of the cloud. If the 
shear viscosity is proportional to the pressure then the system 
freezes out at some time $t_{\it fr}$. For values of $\eta/P$ 
implied by the data this time is much larger than the characteristic
time for dissipative effects in the evolution of the system, and
the estimates in Sect.~\ref{sec_diss_de}-\ref{sec_diss_ns} are 
internally consistent. If the shear viscosity is constant then
there is a freezeout surface which moves inward as a function
of time. This implies that the integral in equ.~(\ref{e_dot}) and
(\ref{ns_mom}) should be restricted to the region enclosed by 
the freezeout surface. However, in order for energy to be conserved, 
and for viscosity to have an effect on the evolution of the 
system, we would have to include an external force on the freezeout
surface. 

 An approach that can describe the effects of freezeout without 
the need to introduce an artificial surface is second order viscous
hydrodynamic \cite{Garcia:2008}. The second order formalism takes
into account terms with two derivatives of the thermodynamic variables 
in the dissipative correction to the stress tensor and energy current. 
In general, the second order formalism contains a large number of new 
transport coefficients. A phenomenological ansatz that has proven to be 
useful in many different applications is to treat the viscous part of 
the stress tensor as an independent hydrodynamical variable which 
satisfies a relaxation equation
\be 
\label{del_pi_is}
\tau_R\frac{\partial}{\partial t}  \delta\Pi_{ij} =- 
   \delta\Pi_{ij}+\delta\Pi^{NS}_{ij}\, ,
\ee
where $\tau_R$ is the relaxation time and $\delta\Pi^{NS}_{ij}$
is the Navier-Stokes expression for the viscous contribution to the
stress tensor, equ.~(\ref{del_pi_ns}). An equation of this type
was first introduced by Maxwell and Cattaneo in the context of heat 
transport. More recently, time or frequency dependent viscosities
were considered in the study of Bose condensed gases in 
\cite{Nikuni:2004,Griffin:2009}. In relativistic hydrodynamics 
relaxation equations for the viscous stress tensor are used in 
order to restore causality, see the review \cite{Romatschke:2009im}. 

 Scale invariance implies that $\tau_R(n,T)=w(mT/n^{2/3})/T$ where 
$w(y)$ is a universal function. In the dilute limit $y\gg 1$ the 
function $w(y)$ can be calculated in kinetic theory which gives 
$\tau_R=\eta/(nT)$ \cite{Bruun:2007}. This result corresponds to 
the estimate for $\tau_{\it mft}$ given in equ.~(\ref{fr_out}).
The relaxation equation (\ref{del_pi_is}) requires an initial condition 
for the viscous stress $\delta\Pi_{ij}$. If is natural to assume that 
$\delta\Pi_{ij}=0$ at $t=0$. In the center of the cloud $\tau_R$
is small and the viscous stress quickly relaxes to the Navier-Stokes
result.  In the dilute region $\tau_R\to\infty$ and the viscous
contribution to the stress tensor remains zero. This implies that 
even a spatially constant shear viscosity leads to a spatially 
varying $\delta\Pi_{ij}$ and a non-zero drag force. This drag 
force is largest near the freezeout surface and breaks the scaling
nature of the flow. This means that a detailed study of the 
Israel-Stewart equations will require numerical solutions of the 
hydrodynamic equations. We can estimate the effect of the 
relaxation time by computing the energy dissipation. We have
\be 
\label{del_e_is}
\dot E = -\frac{1}{2}\int d^3x\, \delta\Pi_{ij}
 \left( \nabla_i v_j+\nabla_j v_i
  - \frac{2}{3}\delta_{ij}\nabla\cdot v\right),  
\ee 
where $\delta\Pi_{ij}$ is determined by equ.~(\ref{del_pi_is}). The 
simplest approximation is to set $\delta\Pi_{ij}=\delta\Pi_{ij}^{NS}$
inside the freezeout surface and $\delta\Pi_{ij}=0$ outside. 

\begin{figure}[t]
\bc\includegraphics[width=8.5cm]{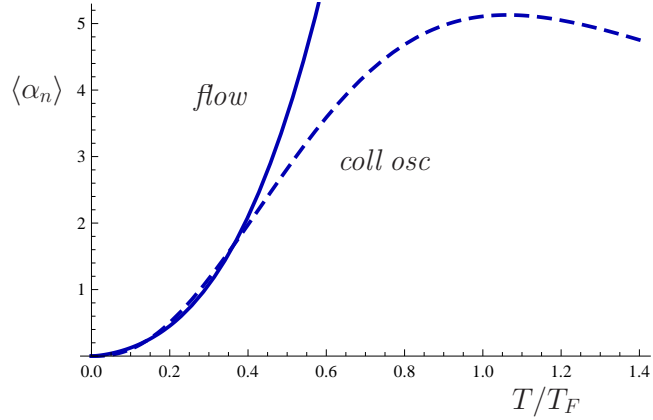}\ec
\caption{\label{fig_relax}
Trap average $\langle \alpha_n\rangle=\langle\eta\rangle/N$
computed from a relaxation time equation with $\eta=\eta_0(mT)^{3/2}$
and $\tau_R=\eta/(nT)$. Contrary to the pure Navier-Stokes case 
$\tau_R\to 0$ the ratio $\langle\eta\rangle/N$ depends on the number 
of particles and the trap geometry. Here we have chosen $N=2\cdot 10^5$ 
and $\lambda=0.045$. The solid shows the result for the elliptic 
flow field, and the dashed line corresponds to the transverse collective 
mode, see Sect.~\ref{sec_coll}. }
\end{figure}

 In order to obtain more accurate estimates we have to solve the 
differential equation (\ref{del_pi_is}). As in Sect.~\ref{sec_diss_de}
we may compute $\delta\Pi_{ij}^{NS}$ from the solution of ideal 
hydrodynamics. The relaxation time can be calculated using the 
high temperature result for the density profile. We find
\be
\omega_\perp\tau_R = \frac{45\pi}{8\sqrt{2}}
  \frac{1}{(3\lambda N)^{1/3}}
  \left(\frac{T}{T_F}\right)^2  b_\perp^{4/3}  
  \exp\left(\frac{x_\perp^2}{b_\perp^2 {\bar R}^2_\perp}
          +\frac{x_z^2}{{\bar R}_z^2}\right)\, ,
\ee
which has the same functional form as the freezeout criterion in 
equ.~(\ref{fr_out_eta_0}). The viscous stress tensor $\delta\Pi_{ij}$ 
is determined by integrating equ.~(\ref{del_pi_is}) and the dissipated 
energy can be computed from equ.~(\ref{del_e_is}). By comparing 
$\Delta E$ with equ.~(\ref{e_dot}) we can express the result  
in terms of an effective $\langle\alpha_n\rangle$. This quantity 
is shown in Fig.~\ref{fig_relax}. We observe that $\langle\alpha_n
\rangle$ grows with temperature as $\langle\alpha_n \rangle\sim T^3$,
much faster than one would expect from the relation $\eta\sim T^{3/2}$. 

 There are no data for elliptic flow at temperatures above $T_c$, but 
we will compare the relaxation time result to collective mode data 
in Sect.~\ref{sec_coll}. We note that at low temperature the 
effective $\langle \alpha_n\rangle$ is the same for expanding and 
oscillating systems, but that at high temperature the two systems
behave differently. In the expanding system the hydrodynamic 
expansion time $\tau_{\it exp}$ continues to increase during the 
expansion, whereas the period of the oscillation provides a fixed 
hydrodynamic time scale in the case of the collective mode. The 
viscous relaxation time $\tau_R$ increases with temperature. This 
implies that for the collective mode we eventually get $\tau_R>
\tau_{\it exp}$ and the effective $\langle\alpha_n\rangle$ starts 
to decrease. In the expanding system, on the other hand, the 
relaxation time can always match the expansion time and $\langle
\alpha_n\rangle$ continues to grow with temperature.

\section{Expansion from a rotating trap}
\label{sec_rot}

 The expansion from a rotating trap was studied in \cite{Clancy:2007}.
Rotating gases are of interest for a number of reasons. The quenching
of the moment of inertia in a superfluid Bose gas was used as a 
signature of superfluidity \cite{Edwards:2002}. The remarkable discovery 
in \cite{Clancy:2007} is that in a Fermi gas at unitarity the suppression 
of the moment of inertia is also observed in the normal phase. It is 
clearly of interest to determine to what extent this discovery places 
constraints on the shear viscosity \cite{Thomas:2009}.

\subsection{Ideal fluid dynamics}
\label{sec_rot_id}

 The Euler equations for a Bose gas with $P\sim n$ were derived in 
\cite{Edwards:2002}. The result is easily generalized to a Fermi 
gas at unitarity \cite{Clancy:2007}. As in the case of a non-rotating 
trap the equations are independent of the temperature and the universal 
function $f(y)$ in equ.~(\ref{P_uni}). We have
\bea
\label{euler_rot_1}
\dot \alpha_x + \alpha_x^2 + \alpha^2 - \Omega^2 &=& 
  \frac{\bar{\mu}\omega_x^2}{b_x^2} \\
\dot \alpha_y + \alpha_y^2 + \alpha^2 - \Omega^2 &=& 
  \frac{\bar{\mu}\omega_y^2}{b_y^2} \\
\dot \alpha_z + \alpha_z^2  \;\;\;\; &=& 
  \frac{\bar{\mu}\omega_z^2}{b_z^2}  \\
\dot \alpha + \alpha \left(\alpha_x+\alpha_y\right) &=& 
 \frac{\bar{\mu} a \omega_x^2}{2}  \\[0.1cm]
\label{euler_rot_5}
\dot \Omega + \Omega \left(\alpha_x+\alpha_y \right) &=& 0 \, . 
\eea
These equations have to be solved together with the continuity 
equations (\ref{cont_1}-\ref{cont_5}). In all there are ten 
coupled equations. In the case of a rotating trap there is no 
initial expansion, $\alpha_i(0)=0$, but either $\alpha(0)$ or 
$\Omega(0)$ (or both) are non-zero. If the initial flow is purely 
irrotational then $\alpha(0)=\omega_{\it rot}$, where $\omega_{\it rot}$ 
is the angular velocity of the trap. If the flow corresponds
to rigid rotation then $\Omega(0)=\omega_{\it rot}$. Below 
the critical temperature the flow of the superfluid component 
must be irrotational, but above $T_c$ both rotational and 
irrotational flows are possible. 

\begin{figure}[t]
\bc\includegraphics[width=10cm]{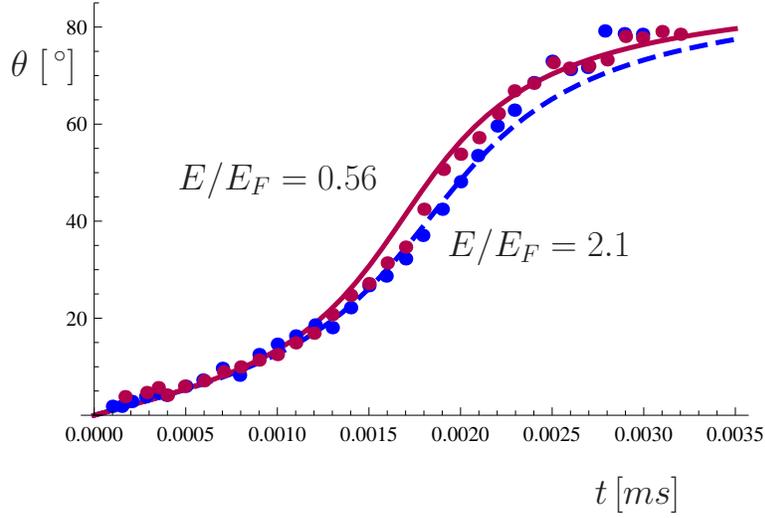}\ec
\caption{\label{fig_rot}
Time evolution of the angle of the major axis of a rotating expanding 
cloud after release from the trapping potential. The data are taken from 
\cite{Clancy:2007}. The two data sets were obtained with initial energies 
$E/E_F=0.56$ and 2.1. The solid line shows the prediction of ideal 
fluid dynamics, and the dashed lines shows the solution of the 
Navier-Stokes equation for $\beta=0.061$. Using an entropy per particle 
$S/N\simeq 4.8$ this value of $\beta$ implies a shear viscosity to 
entropy density ratio $\langle\alpha_s\rangle=0.60$ }
\end{figure}

 The equations simplify in the experimentally relevant case 
of strongly deformed, slowly rotating traps, $\omega_{rot}<
\omega_x\ll \omega_\perp$ with $\omega_\perp=\omega_y\simeq \omega_z$. 
In this limit the motion of the fluid is dominated by the transverse 
expansion of the system. Up to corrections of order $O(\lambda^2)$ or 
$O((\omega_{\it rot}/\omega_\perp)^2)$ we have 
\be 
 b_\perp(t) \simeq\left\{ \begin{array}{cl}
 1+\frac{1}{2}\omega_\perp^2t^2 + O(t^4)\;\;\;\; 
                        & \omega_\perp t\ll 1 \, , \\
 \frac{\omega_\perp t}{\sqrt{\gamma}}  + c_0 + O(t^{-1/3}) 
                        & \omega_\perp t\gg 1 \, , 
\end{array}\right.
\ee
as in the case of a stationary trap. The orientation of the 
expanding cloud is described by the parameter $a$ defined 
in equ.~(\ref{scal_par}). We find
\be 
\label{a_as}
a(t)  \simeq \left\{ \begin{array}{cl}
  -\frac{2\omega_{\it rot}t}{\lambda^2} \;\; 
                    & \omega_\perp t\ll 1 \, , \\
  - \frac{c_a\omega_{\it rot}}{\lambda^2\omega_\perp^2 t}
                     & \omega_\perp t\gg 1 \;\; (t<t_{3d})  ,
\end{array}\right.
\ee
where $c_a$ is a constant. Below we will show that $c_a=\gamma$. 
At very late times, $t>t_{3d}\sim 1/(\lambda^2\omega_\perp)$, 
we find $a(t)\sim 1/t^2$. The result (\ref{a_as}) holds 
irrespective of the nature of the initial rotational flow.  
The parameter $a(t)$ can be related to the angle of the cloud
with respect to the $x$-axis, 
\be
\tan(2\theta) = - \frac{a\lambda^2b_x^2b_y^2}{b_x^2-\lambda^2 b_y^2}\, . 
\ee
At early times, $\omega_x t\ll 1$, the angle is proportional 
to the rotational frequency of the trap, $\theta = \omega_{\it rot}t$. 
The angular motion speeds up as $b_y\lambda$ approaches $b_x$. The 
angle goes through $45^\circ$ at
\be 
 t_{45^\circ} = \frac{\sqrt{\gamma}}{\omega_x}\, 
\ee
\begin{figure}[t]
\bc\includegraphics[width=9cm]{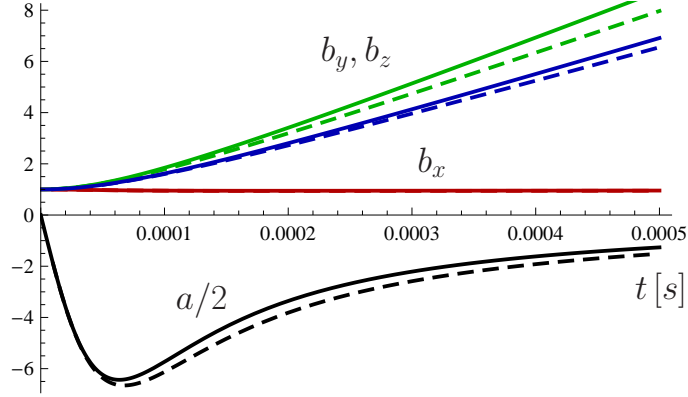}\ec
\caption{\label{fig_rot_flow}
Time evolution of the parameters $a,b_x,b_y,b_z$ that characterize 
the scaling expansion out of a rotating trap. 
Note that in this case $\omega_y$ 
and $\omega_z$ are not exactly equal, and that the time scale is 
different from Fig.~\ref{fig_rot}. Here, we only show the early 
evolution of the system. Solid lines show the solution of the Euler 
equation, and dashed lines show the solution of the Navier-Stokes 
equation for $\beta=0.077$. }
\end{figure}
which is the identical to the crossing time in equ.~(\ref{t_cross}). At 
late times, and up to corrections of $O(\omega_{\it rot}/\omega_\perp)$, 
the angle approaches $90^\circ$. The velocity field is dominated by 
the transverse expansion of the system. In the limit $\omega_{rot}<
\omega_x\ll \omega_\perp$ the velocity fields $\alpha_i$ are identical 
to those in the non-rotating case. We have
\be 
 \alpha_{y,z} \simeq\left\{ \begin{array}{cl}
  \omega_\perp^2t \;\; & \omega_\perp t\ll 1 \, , \\
   1/t  & \omega_\perp t\gg 1 \, , 
\end{array}\right.
\ee
and $\alpha_x=O(\lambda^2)$. The rotational components of the 
velocity field decay quickly. If the initial flow is irrotational, 
$\alpha(0)=\omega_{\it rot}$, then 
\be 
\alpha(t) \simeq \omega_{\it rot} \left( 1- \omega_\perp^2 t^2\right)\, 
\ee
for $(\omega_\perp t)<1$. For $(\omega_\perp t)>1$ the rotational
component of the flow is small, $(\alpha/\omega_{\it rot})\ll 1$, 
but the remaining flow decays slowly, $\alpha\sim t^{-1}$ for 
$t<t_{3d}$ and $\alpha\sim t^{-2}$ for $t>t_{3d}$. In ideal 
hydrodynamics an initially irrotational flow will remain irrotational, 
$\Omega(t)=0$, for all $t$. If the initial flow corresponds to rigid 
rotation, $\Omega(0)=\omega_{\it rot}$, then the early time behavior 
is given by
\be 
\Omega(t) \simeq \omega_{\it rot} 
  \left( 1- \frac{1}{2}\omega_\perp^2 t^2\right)\, . 
\ee
An initially rigid rotating flow induces a non-zero irrotational 
flow. For $(\omega_\perp t)>1$ both components of the velocity 
field become much smaller than $\omega_{\it rot}$. 

\begin{figure}[t]
\bc\includegraphics[width=9cm]{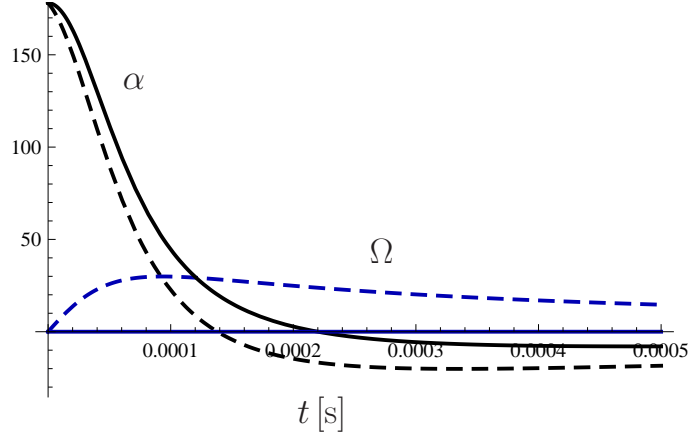}\ec
\caption{\label{fig_rot_flow_2}
Time evolution of the parameters $\alpha$ and $\Omega$ which control
the irrotational and rotational components of the velocity field.
Parameters were chosen as in Fig.~\ref{fig_rot_flow}. Solid lines show 
the solution of the Euler equation, and dashed lines show 
the solution of the Navier-Stokes equation for $\beta=0.077$. }
\end{figure}

The angular momentum is given by
\be 
\label{l_z}
 L_z = \alpha m\langle n(x^2-y^2)\rangle 
      +\Omega m\langle n(x^2+y^2)\rangle 
      +\left(\alpha_x-\alpha_y\right)m\langle nxy \rangle ,
\ee
where $n$ is the density and $\langle .\rangle$ is an integral over 
the cloud. The moment of inertia of a rigid rotor is $I_{\it rig}=
m\langle x^2+y^2\rangle$, and the irrotational moment of inertia 
is $I_{\it irr}=m\langle x^2-y^2\rangle$. We have 
\bea 
 m\langle n x^2\rangle &=& 
  \frac{b_x^2}{1-\frac{\lambda^2}{4}(ab_xb_y)^2}
  \,\frac{L_0}{\omega_x}\, ,  \\
 m\langle n y^2\rangle &=& 
  \frac{\lambda^2b_y^2}{1-\frac{\lambda^2}{4}(ab_xb_y)^2}
  \,\frac{L_0}{\omega_x}\, ,  \\
 m\langle n xy\rangle &=& 
  \frac{-\frac{\lambda^2}{2}ab_x^2b_y^2}{1-\frac{\lambda^2}{4}(ab_xb_y)^2}
  \,\frac{L_0}{\omega_x} \, , 
\eea
where the scale is set by 
\be 
 L_0 = \frac{N}{6} \frac{(3N)^{1/3}}{\lambda^{2/3}} 
   \left( \frac{E_0}{E_F}\right)\, . 
\ee
In the experiment of Clancy et al.~$(\omega_{\it rot}/\omega_x)\simeq
0.4$ and $L_0/N\simeq 131 (E_0/E_F)$. For $E_0/E_F=1$, which is in the
normal phase, the angular momentum per particle is 50$\hbar$.

 At early times the trap is strongly deformed and $I_{\it rig}\simeq 
I_{\it irr}$. When the cloud becomes almost spherical the irrotational 
moment is much smaller than the rigid moment of inertia, $I_{\it irr}
\ll I_{\it rig}$. However, at times $(\omega_\perp t)>1$ the angular 
momentum is mainly carried by the last term in equ.~(\ref{l_z}), 
which is related to the transverse expansion of the system. This 
is true irrespective of the nature of the initial rotational flow.
For $(\omega_\perp t)>1$ we have $\alpha_y m\langle nxy\rangle\simeq 
(c_a/\gamma)(\omega_{\it rot}/\omega_x)L_0$. Angular momentum 
conservation then fixes the constant $c_a$ in equ.~(\ref{a_as}), 
$c_a=\gamma$. At very late time, $t>t_{3d}$, the angular momentum 
is shared among all the terms in equ.~(\ref{l_z}), and the 
relative size of the different contributions depends on the 
initial conditions. In practice, of course, hydrodynamics is 
no longer applicable at $t>t_{3d}$. 

\begin{figure}[t]
\bc\includegraphics[width=10cm]{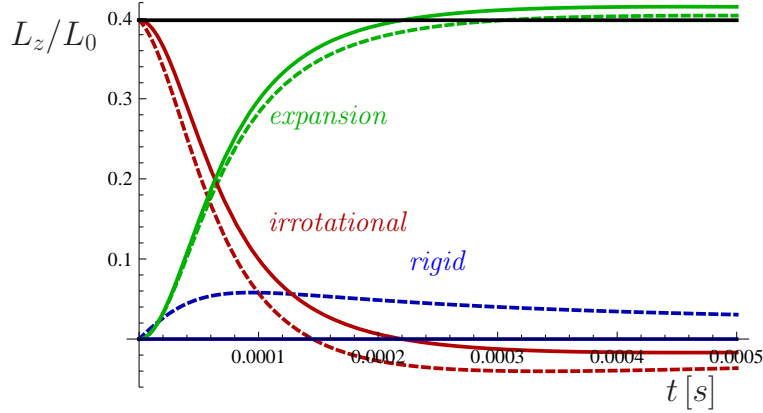}\ec
\caption{\label{fig_Lz}
This figure shows different contribution to the total angular 
momentum of the expanding cloud as a function of time. The angular 
momentum is given in units of the quantity $L_0$ defined in the 
text. The curves labeled irrotational, rigid, and expansion show 
the $\langle x^2-y^2\rangle$, $\langle x^2+y^2\rangle$, and $\langle xy
\rangle$ contributions. The solid and dashed lines correspond
to ideal and viscous hydrodynamics, respectively. The solid black
line shows the (conserved) total angular momentum.    }
\end{figure}

\subsection{Dissipation}
\label{sec_rot_diss}

 The effects of dissipation on the expansion from a rotating 
trap can be studied in close analogy with 
Sect.~\ref{sec_diss_de}-\ref{sec_diss_relax}. The rate of energy 
dissipation is
\be 
 \dot E = -\frac{4}{3} \left( \alpha_x^2 + \alpha_y^2 + \alpha_z^2
  - \alpha_x\alpha_y - \alpha_x\alpha_z -\alpha_y\alpha_z
  + 3\alpha^2 \right) \, 
  \int d^3x\, \eta(x)\, .
\ee
For $\alpha_x\simeq \alpha_y\gg \alpha_z,\alpha$ this expression
reduces to the energy dissipated by the transverse expansion of 
cloud, see equ.~(\ref{e_dot}). This implies that the main effect
of dissipation is to slow the transverse expansion of the cloud, 
and to delay the time $t_{45^\circ}$. This delay is exactly the 
same as the delay in the crossing time in equ.~(\ref{del_t_cr}).
We have
\be 
\label{del_t_rot}
 \left(\frac{\delta t}{t}\right)_{45^\circ} 
 = 0.009
  \left( \frac{\langle \alpha_s\rangle}{1/(4\pi)}\right)\,
  \left( \frac{1.3\cdot 10^5}{N} \right)^{1/3}
  \left( \frac{0.3}{\lambda} \right)^{1/3}
  \left( \frac{S/N}{4.8} \right)\,
  \left( \frac{2.1}{E_0/E_F} \right)\, . 
\ee
We can confirm this estimate by solving the Navier-Stokes equation.
The Navier-Stokes equation can be derived using the moment method 
described in Sect.~\ref{sec_diss_mom}. As before, an equivalent 
set of equations can be obtained from the viscosity model given in 
equ.~(\ref{eta_scal}). We get \cite{Clancy:2008}
\bea
\label{ns_rot_1}
\dot \alpha_x + \alpha_x^2 + \alpha^2 - \Omega^2 &=& 
  \frac{\omega_x^2}{b_x^2}
  \left\{ \bar{\mu} - \frac{6\beta}{\omega_\perp}\left[
    \frac{2}{3}\alpha_x-\frac{1}{3}\left(\alpha_y+\alpha_z\right)
   + \frac{1}{2} ab_x^2\alpha \right]\right\}
 \\
\dot \alpha_y + \alpha_y^2 + \alpha^2 - \Omega^2 &=& 
  \frac{\omega_y^2}{b_y^2} 
   \left\{ \bar{\mu} - \frac{6\beta}{\omega_\perp}\left[
    \frac{2}{3}\alpha_y-\frac{1}{3}\left(\alpha_x+\alpha_z\right)
   + \frac{1}{2} a\lambda^2b_y^2\alpha \right]\right\}
\\
\dot \alpha_z + \alpha_z^2  \;\;\;\; &=& 
  \frac{\omega_z^2}{b_z^2}  
  \left\{ \bar{\mu} - \frac{6\beta}{\omega_\perp}\left[
    \frac{2}{3}\alpha_z-\frac{1}{3}\left(\alpha_x+\alpha_y\right)
      \right]\right\}
\\
\dot \alpha + \alpha \left(\alpha_x+\alpha_y\right) &=& 
  \omega_x^2
   \left\{ \frac{\bar{\mu}a}{2}   - \frac{3\beta}{\omega_\perp} \left[
       \frac{a}{6}\left(\alpha_x+\alpha_y-2\alpha_z\right)
   + \frac{b_x^2+\lambda^2b_y^2}{\lambda^2b_x^2b_y^2}\,\alpha\right]\!\!
  \right\}
\\[0.1cm]
\label{ns_rot_5}
\dot \Omega + \Omega \left(\alpha_x+\alpha_y \right) &=&   
  \frac{3\beta \omega_x^2}{\omega_\perp}   \left[
       \frac{a}{2}\left(\alpha_x-\alpha_y\right)
   + \frac{b_x^2-\lambda^2b_y^2}{\lambda^2b_x^2b_y^2}\,\alpha\right]
 \, . 
\eea
\begin{figure}[t]
\bc\includegraphics[width=9cm]{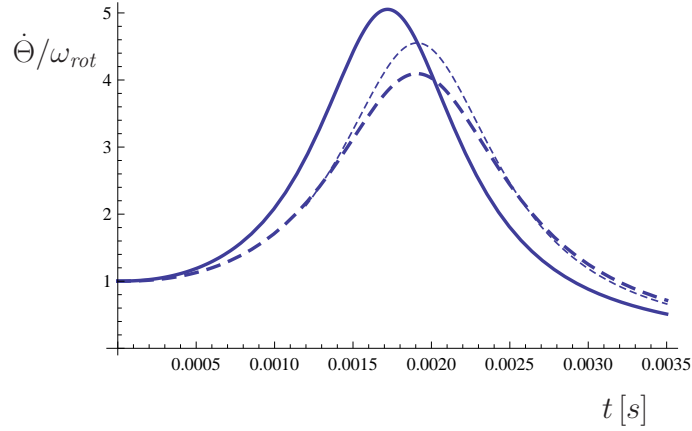}\ec
\caption{\label{fig_theta_dot}
This figure shows the angular velocity of the rotating cloud
as a function of time. The solid line shows the solution of the
Euler equation, and the dashed line is the solution of the 
Navier-Stokes equation for $\beta=0.077$. The thin dashed line 
shows the result for the angular velocity obtained by rescaling the 
solution of the Euler equation by a factor $1+(\delta t/t)_{45^\circ}
\simeq 1.1$. The discrepancy between the Navier-Stokes prediction and 
the rescaled Euler result in the regime where $\dot\Theta$ is 
large is due to the rotational component of the flow. We note 
that $I=L/\dot\Theta$ is the moment of inertia. }
\end{figure}
These equations are independent of the functional form of the 
pressure. A solution of the Navier-Stokes equation for the trap 
parameters and initial conditions in \cite{Clancy:2007} is shown 
in Fig.~\ref{fig_rot}. The experimental data were taken at $E/E_F=0.56$ 
which is in the superfluid phase, and $E/E_F=2.1$ which is significantly 
above the phase transition. Similar to the low temperature data 
for pure transverse expansion in Fig.~\ref{fig_flow} the low 
temperature result for a rotating cloud shows no dissipative effects, 
and the best fit to the data is provided by ideal fluid dynamics.

 The data for $E/E_F=2.1$ clearly show a delayed expansion. We find 
$(\delta t/t)_{45^\circ}\simeq 0.063$. Using $(\delta t/t)\simeq 1.16
\beta$ from equ.~(\ref{delE_lin_beta}) we estimate $\beta\simeq 0.057$.
 This estimate is quite accurate, the best fit of the Navier-Stokes 
solution to the data is obtained for $\beta=0.061$. Using $N=1.3
\cdot 10^5$, $\lambda=0.03$ \cite{Clancy:2007} and $(S/N)\simeq 4.8$ 
\cite{Luo:2008} we obtain $\langle \alpha_s\rangle\simeq 0.60$. The 
measurements were extended to values of $E/E_F$ between 0.56 and 2.1 
in \cite{Thomas:2009}. This work reports values of $\eta/s$ as small 
as $\langle \alpha_s\rangle\simeq (0.0-0.4)$. Note that in this regime 
it becomes very difficult to measure the viscosity accurately. A
value of $\langle \alpha_s\rangle=0.1$ affects the measured angle 
of the cloud by less than the with of the lines in Fig.~\ref{fig_rot}.

 A more detailed study of viscous effects on the evolution of the 
system is shown in Figs.~\ref{fig_rot_flow}-\ref{fig_Lz}. We observe
that viscosity slows down the evolution of the scale parameters 
$b_y,b_z$ and $a$. More interesting is the effect on the velocity
fields $\alpha$ and $\Omega$. Viscosity converts a fraction of the 
irrotational velocity field $\alpha$ into the rotational velocity 
field $\Omega$. This is also seen in the breakdown of the angular 
momentum, see Fig.~\ref{fig_Lz}. The rotational component of $L_z$
is not large, but it does lead to an observable effect in the 
angular velocity of the cloud. Fig.~\ref{fig_theta_dot} shows that 
viscosity leads to a decrease in $\dot\Theta$. During most of the 
evolution this effect is dominated by the delayed expansion, but 
for $t\simeq t_{45\circ}$ there is an extra reduction which is due 
to an increase of the effective moment of inertia $I=L/\dot\Theta$
caused by the rotational flow. Unfortunately, the experimental data 
are for $\Theta(t)$ are not sufficiently accurate to demonstrate
this effect.

\begin{figure}[t]
\bc\includegraphics[width=10cm]{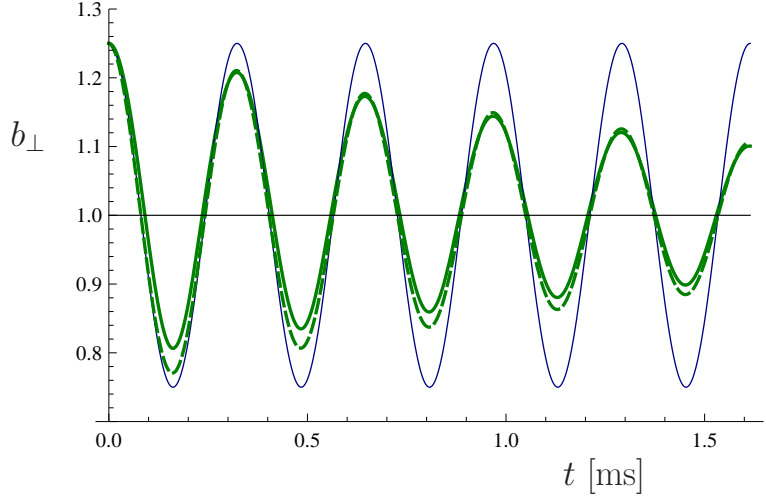}\ec
\caption{\label{fig_osc}
Time evolution of the amplitude of the transverse breathing
mode. The black line shows the solution of the Euler equation and 
the solid green line is the solution of the Navier-Stokes equation 
for $\beta=0.05$. The dashed green line is the damped cosine function
given in equ.~(\ref{damp_cos}). The trap frequency was chosen to be 
$\omega_\perp=1696$ Hz as in \cite{Kinast:2005}.}
\end{figure}

\section{Collective Oscillations}
\label{sec_coll}

 In order to study collective oscillations we consider the 
Euler equation (\ref{euler_flow}) in the presence of the trapping 
potential. The equation of motion is 
\be 
\label{euler_trap}
 \ddot b_i = \frac{\omega_i^2}{(b_xb_yb_z)^{2/3}}\frac{1}{b_i}
  - \omega_i^2 b_i \, . 
\ee
The equilibrium solution is $b_x=b_y=b_z=1$. We now consider 
small oscillations around the equilibrium, $b_i(t)=1+ a_i
e^{i\omega t}$. The linearized equation of motion gives
\be 
\omega^2 a_i = \omega_i^2 \left(2a_i + \gamma \sum_j a_j \right),
\ee
which was derived in \cite{Heiselberg:2004,Stringari:2004,Bulgac:2004}
using slightly different methods. For the radial breathing mode 
$a_y=a_z=a_\perp$, $a_x=0$ we get $\omega^2=2(1+\gamma)\omega_\perp^2
=(10/3)\omega_\perp^2$. The energy dissipated can be computed from
equ.~(\ref{e_dot}). We find
\be 
\label{del_e_osc}
 \frac{\Delta E}{E_{\it osc}} = -4\pi \sqrt{\frac{3}{10}}\, \beta
 \simeq -6.88\cdot\beta\, , 
\ee
where $\Delta E$ is the energy dissipated per period, $E_{\it osc}$ 
is the energy of the collective mode, and $\beta$ is the parameter 
defined in equ.~(\ref{beta}). We note that the amount of energy 
dissipated in one period of the transverse breathing mode is 
about three times larger than the energy dissipated by transverse
expansion, see equ.~(\ref{delE_lin_beta}).

We can also derive a Navier-Stokes equation, either by taking moments 
as in Sect.~\ref{sec_diss_mom}, or by using a simple scaling form of 
the shear viscosity as in Sect.~\ref{sec_diss_ns}. For the 
transverse breathing mode we find
\be 
\label{ns_trap}
 \ddot b_\perp = \frac{\omega_\perp^2}{b_\perp^{7/3}}
  - \omega_\perp^2b_\perp
  - \frac{2\beta\omega_\perp\dot{b}_\perp}{b_\perp^2}\, . 
\ee
If $\beta$ is small then this equation is approximately solved 
by a damped oscillating function. We have
\be 
\label{damp_cos}
 b_\perp(t) = 1 + a_\perp \cos(\omega t) \exp(-\Gamma t)\, . 
\ee
Comparison with equ.~(\ref{del_e_osc}) gives $\Gamma=\beta\omega_\perp$.
The main feature of collective modes is that the viscous term exponentiates 
so that even very small values of $\beta$ are experimentally accessible. 
In Fig.~(\ref{fig_osc}) we show a comparison between an exact solution 
of equ.~(\ref{ns_trap}) for $\beta=0.05$, $a_\perp(0)=0.25$ and 
the approximate solution (\ref{damp_cos}). We observe that the 
approximate solution is extremely accurate. 

\begin{figure}[t]
\bc\includegraphics[width=10cm]{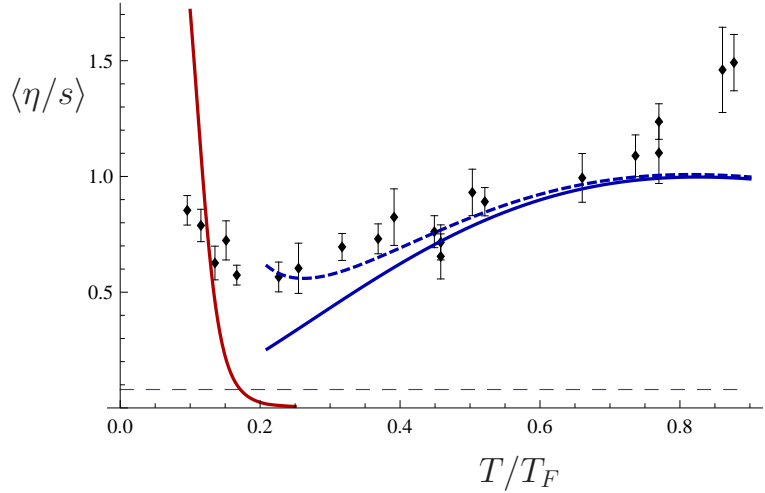}\ec
\caption{\label{fig_eta_s_osc}
Trap average $\langle\alpha_s\rangle = \langle \eta/s\rangle$ 
extracted from the damping of the radial breathing mode. The
data points were obtained using equ.~(\ref{alpha_s_data}) to 
analyze the data published by Kinast et al.~\cite{Kinast:2004b}.
The thermodynamic quantities $(S/N)$ and $E_0/E_F$ were taken
from \cite{Luo:2008}. The solid red and blue lines show the 
expected low and high temperature limits. Both theory curves
include relaxation time effects. The blue dashed curve is 
a phenomenological two-component model explained in the text. }
\end{figure}

 The experimentally measured damping rate can be used to estimate 
$\langle\alpha_s\rangle$. We have 
\be 
\label{alpha_s_data}
\langle \alpha_s \rangle = (3\lambda N)^{1/3}
   \left(\frac{\Gamma}{\omega_\perp}\right) 
   \left(\frac{E_0}{E_F}\right)\left(\frac{N}{S}\right)\, . 
\ee
In Fig.~\ref{fig_eta_s_osc} we show an analysis of the data obtained 
by Kinast et al.~\cite{Kinast:2004b} using equ.~(\ref{alpha_s_data}). 
This plot is very similar to our earlier analysis \cite{Schafer:2007pr} 
(see also \cite{Gelman:2004,Turlapov:2007}), except that the 
temperature calibration and thermodynamic data have been updated
using the recent analysis published in \cite{Luo:2008}.
 
There are a number of important checks on the interpretation
of the damping date in terms of viscous hydrodynamics that should 
be, or have already been, performed. Viscous hydrodynamics predicts 
that the monopole mode in a spherical trap is not damped at all. 
This prediction is quite striking, but it has never been tested. 
Viscous hydrodynamics also predicts simple relationships between 
the damping constant of the radial breathing mode and the radial 
quadrupole as well as the scissors mode \cite{Schafer:2007pr}. These 
predictions agree qualitatively with the data obtained by the Innsbruck 
group, but there are some structures in the data that do not fit a
simple hydrodynamic description. Finally, hydrodynamics predicts 
that the damping rate decreases as $N^{-1/3}$. This prediction does 
not agree with the data published in \cite{Kinast:2004b}. We note, 
however, that Kinast et al.~only checked the scaling behavior at 
very low temperature, and that relaxation time effects may modify 
the particle number scaling. 

\begin{figure}[t]
\bc\includegraphics[width=8.5cm]{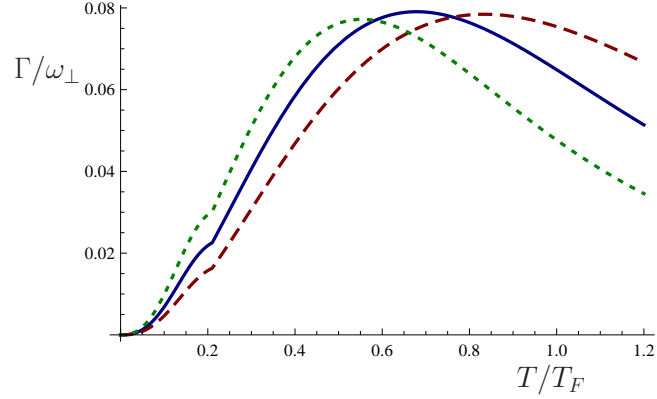}\ec
\caption{\label{fig_damp_rel}
Damping rate of the radial breathing mode in units of the transverse 
trapping frequency. This figure only shows the contribution from
the dilute corona, computed using the relaxation time approach. 
The solid line corresponds to $N\equiv N_0=2\cdot 10^5$, $\lambda=0.045$
as in \cite{Kinast:2004b}. The long dashed and short dashed lines
corresponds to $N=5N_0$ and $N=0.2 N_0$, respectively.}
\end{figure}

 We can also compare the results in Fig.~\ref{fig_eta_s_osc}
to theoretical prediction for the shear viscosity in the 
low and high temperature limit. In the high temperature 
limit the viscosity is independent of density and the 
main source of dissipation is the finite relaxation time, 
see Sec.~\ref{sec_diss_relax}. In the case of periodic
motion the relaxation time equation (\ref{del_pi_is}) is 
easily solved. The dissipated energy is given by 
equ.~(\ref{beta},\ref{del_e_osc}) with 
\be 
\langle \alpha_n\rangle = \eta_0 (mT)^{3/2} 
 \int d^3x\, \frac{1}{1+\omega^2\tau_R(n(x))^2}. 
\ee
We will use the kinetic theory result $\tau_R(n)=\eta/(nT)$ 
with $\eta=\eta_0(mT)^{3/2}$. In the high temperature (low
density) limit we can use the classical expression for the 
density profile $n(x)$. In this case the integral over $x$ 
can be done analytically. We find
\be 
\label{alpha_n_relax}
\langle\alpha_n\rangle  = -\frac{45\pi}{32}
 \left(\frac{T}{T_F}\right)^3 {\it Li}_{3/2}
\left( -\left[\frac{{\it const}}{(\lambda N)^{2/3}}
        \left(\frac{T}{T_F}\right)^4\right]^{-1}\right)\, ,
\ee
where ${\it const}=1125\cdot 3^{1/3}\pi^2/64\simeq 250.1$, 
and ${\it Li}_\alpha(x)$ is the polylogarithm function.  
In the limit $T\ll T_F$ the result scales as $\langle 
\alpha_n\rangle \sim y^3\log(y)^{3/2}$ with $y=T/T_F$. 
For $T\gg T_F$ we get $\langle \alpha_n\rangle \sim y^{-1}$. 
These results imply that both the temperature scaling and
the particle number scaling differ from naive expectations. 
The shear viscosity scales as $\eta\sim T^{3/2}$, but 
$\langle \alpha_n\rangle\sim T^3\log(T)^{3/2}$ at low $T$, 
and $\langle \alpha_n\rangle \sim T^{-1}$ at high $T$. 
Also, the scaling of the damping rate with $N$ is $\Gamma
\sim N^{-1/3}\log(N)^{3/2}$ at low $T$ and $\Gamma\sim 
N^{1/3}$ at high $T$, see Fig.~\ref{fig_damp_rel}. This 
implies that there are temperature regions in which the 
dependence of the damping rate on $N$ is small.

 The prediction of equ.~(\ref{alpha_n_relax}) is shown as the 
solid blue line in Fig.~\ref{fig_eta_s_osc}. We observe that 
the relaxation time model agrees well with the data for $T\sim 
(0.5-0.8)T_F$. For temperature less than $0.5T_F$ the observed
damping rate is bigger than the prediction of the relaxation 
model. At very low temperature the shear viscosity is expected 
to be dominated by the phonon contribution \cite{Rupak:2007vp}
\be 
\label{eta_low_T}
 \eta = 0.018 n \left( \frac{n^{2/3}}{mT}\right)^5\, . 
\ee
At low temperature we can compute the trap average by using 
the zero temperature profile. We find $\langle \alpha_n\rangle
\simeq 1.5\cdot 10^{-5}(T_F/T)^5$. This result becomes large 
for $T/T_F<0.1$. In this regime relaxation time effects are 
important, and $\langle\alpha_n\rangle$ at finite frequency 
goes to zero as $T\to 0$. 

 Neither the low temperature nor the high temperature result 
provide a good description of the data in the regime $T\simeq 
(0.15-0.40)T_F$. The dashed blue line in Fig.~\ref{fig_eta_s_osc}
shows a purely phenomenological fit based on the functional 
form $\eta=\eta_0(mT)^{3/2}+\eta_1 n^{5/3}/(mT)$ with $\eta_0 
=15/(32\sqrt{\pi})\simeq 0.264$ and $\eta_1\simeq 0.06$. In 
this case the minimum value of $\eta/n$ is 0.24 which occurs 
below the phase transition at $mT/n^{2/3}\simeq 0.47$. 

\begin{figure}[t]
\bc\includegraphics[width=10cm]{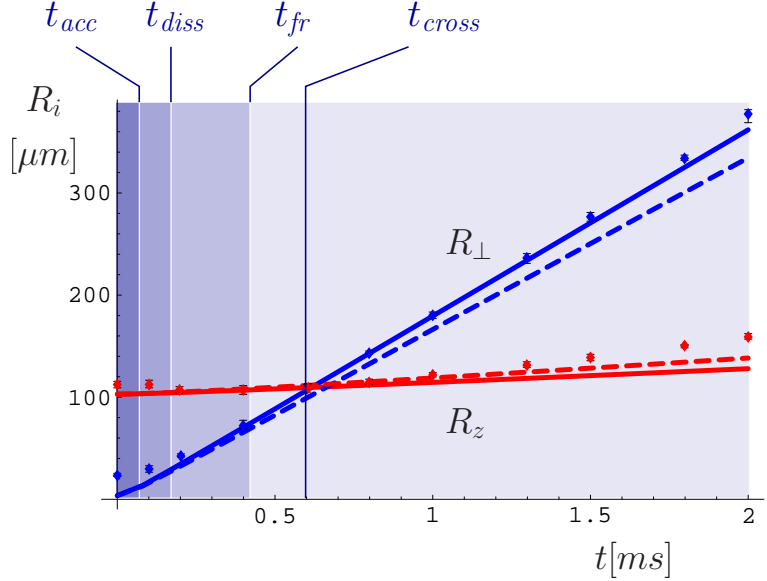}\ec
\caption{\label{fig_scales}
Time scales relevant to the expansion of a unitary Fermi gas
from a deformed trap. The inverse trap frequency is $\omega_\perp^{-1}
=0.024\, ms$. The scale $t_{\it acc}$ is the characteristic time 
for hydrodynamic acceleration, where we have defined $t=t_{\it acc}$ 
to be the time when 80\% of the initial internal energy has been 
converted to kinetic energy. The characteristic time for viscous 
effects, $t_{\it dis}$, is determined by the condition that the 
dissipated energy $\Delta E$ has reached 80\% of its asymptotic 
value. The freezeout time $t_{\it fr}$ is quite uncertain. Here, 
we show the time at which, for $T_0/T_F=0.21$, the freezeout 
surface reaches the point $x_\perp=b_\perp R_\perp$. The crossing 
time $t_{\it cr}$ is the time at which the system becomes spherical. 
The time $t_{3d}$ at which the expansion becomes three-dimensional 
is bigger by another factor $\lambda^{-1}$.   }
\end{figure}

\section{Summary and Outlook}
\label{sec_sum}

 A special feature of the hydrodynamics of a unitary Fermi 
gas is the existence of simple scaling solutions of the 
equations of ideal fluid dynamics. These solutions are 
independent of the equation of state, the initial temperature
and the number of particles. The only time scales in the problem
are the trap frequencies, see Fig.~\ref{fig_scales}. The existence 
of scaling solutions is related to the constraints imposed by 
scale invariance on the equation of state, and to the harmonic 
character of the confinement potential. 

 The properties mentioned above make scaling flows an ideal 
class of solution to study the effects of shear viscosity. In 
this contribution we focused on three classes of experiments, 
expansion from a deformed trap (``elliptic flow''), expansion from 
a rotating trap, and damping of collective oscillations. These 
experiments provide somewhat complementary information, and they 
have different advantages and disadvantages:

\begin{itemize}
\item In the case of collective modes the effect of shear 
viscosity exponentiates, and as a consequence the damping 
of collective modes is sensitive to very small values
of the shear viscosity. Collective modes also have the advantage
that qualitatively the effect of dissipation is very simple: 
The kinetic energy of the collective mode is converted to heat, 
so that at the end of the evolution the system is again stationary, 
but the temperature is increased. In the case of flow experiments 
the situation is more complicated. Dissipation converts kinetic 
energy into heat but unless the system freezes out first, the 
internal energy is eventually converted back to kinetic energy. 
Because of the second law of thermodynamics, the final state 
of viscous hydrodynamics must differ from that of ideal 
hydrodynamics, but the differences can be subtle, manifesting 
themselves in violations of the simple scaling formulas for the 
density and the velocity field. 

\item The transverse expansion experiments provide detailed 
information about the time dependence of the density and flow 
profiles. This information can be used to understand the breakdown 
of hydrodynamics, for example by studying deviations from the 
simple linear velocity profile predicted by ideal fluid dynamics. 
Transverse flow experiments may also show a different, and possibly 
smaller, sensitivity to relaxation effects. Fig.~\ref{fig_relax} 
shows that, for $T/T_F<0.4$, the relaxation time estimate of the 
trap averaged dissipation due to the spatially constant part of 
the shear viscosity is similar for transverse flow and transverse 
collective modes. However, the local response of a rapidly 
expanding cloud is likely to be different from that of an 
oscillating system.

\item The expansion of a rotating cloud is sensitive to a
new viscous effect, the conversion of an irrotational flow 
$\vec{v}\sim \vec{\nabla}(xy)$ to a rotational flow $\vec{v}
\sim \hat{z}\times \vec{x}$. Contrary to the slowdown of the 
transverse expansion, which could in principle be due to 
scale-breaking terms in the pressure or residual external 
potentials, this is a genuine dissipative effect, since 
vorticity is conserved in ideal hydrodynamics. 

\end{itemize}

 The main difficulty in extracting the shear viscosity from 
the analysis of scaling flows is associated with the role of 
the dilute corona of the cloud. Kinetic theory predicts that 
in the dilute limit the shear viscosity is independent of 
density and only depends on temperature. A simple analysis of
the type presented in Sec.~\ref{sec_diss_ns} then implies that 
the dilute corona does not generate a dissipative force. It 
nevertheless dissipates a large amount of energy. The analysis also 
suggests that freezeout only occurs very late, see Sec.~\ref{sec_break}.
There are a number of aspects of this analysis that need to be 
improved:

\begin{itemize}
\item The Navier-Stokes equation is based on the assumption that 
the viscous correction to the stress tensor appears instantaneously.
This is particularly problematic in the case of scaling flows, because 
the viscous contribution is spatially constant. The fact that the 
ideal stresses propagate outward with the expansion of the system
whereas the dissipative stresses appear immediately indicates that
causality is violated. This problem can be addressed by including 
a finite relaxation time, or by solving a more complete set of 
second order hydrodynamic equations.

\item We have studied the effect of dissipative forces in the 
Navier-Stokes equations, but we have computed the non-dissipative
forces (pressure gradients) based on an approximately isentropic
expansion. This procedure neglects reheating, and violates energy 
conservation. Reheating is important in the dilute corona, and
breaks the scaling nature of the expansion. 

\end{itemize}

 In addition to implementing these technical improvements it is 
important to consider other experimental setups that are directly 
sensitive to the spatially constant part of the shear viscosity. One
option would be to measure the attenuation of sound propagating in 
a very long elongated trap. Another idea would be to directly 
measure the decay of a shear flow in a long channel. 

 Finally, we summarize the existing experimental constraints
on the shear viscosity of the unitary Fermi gas:

\begin{itemize}
\item The damping of collective oscillations constrains the 
trap average $\langle\eta\rangle/S\equiv \langle\alpha_s\rangle$. 
We find that this quantity varies between $\langle\alpha_s\rangle
\simeq 1$ at $T/T_F\simeq 0.8$ and $\langle\alpha_s\rangle
\simeq 0.5$ at $T/T_F\simeq 0.2$. In the regime $0.4\leq T/T_F
\leq 0.8$ the temperature dependence is consistent with $\eta
\sim (mT)^{3/2}$ and a relaxation time that scales as $\tau_R
\sim \eta/(nT)$. At lower temperatures an additional contribution
is needed. In a simple model the minimum of the shear viscosity 
to density ratio is $\eta/n\simeq 0.2$. 

\item The expansion of a rotating cloud gives $\langle\alpha_s\rangle
\simeq 0.8$ at $T/T_F\simeq 0.8$, and $\langle\alpha_s\rangle
\simeq (0.0-0.4)$ at $T/T_F\simeq 0.2$ \cite{Thomas:2009}. The 
latter results are smaller than the values extracted from collective
oscillations, although the errors are also somewhat larger. It will
be important to determine whether this discrepancy is due to the 
effects of the dilute corona, and whether the smaller values
of $\langle\alpha_s\rangle$ are more representative of the 
shear viscosity to entropy density ratio in the core. 

\end{itemize}
 
Note added: After the initial version of this contribution
was finished dissipative effects in the expansion of a dilute
Fermi gas at temperatures $T\gg T_F$ were studied experimentally 
by Cao et al.~\cite{Cao:2010wa}. This work nicely demonstrates
the scaling $\langle\alpha_n\rangle \sim T^3$ predicted in 
Fig.~\ref{fig_relax}. Numerical solutions to the equations
of dissipative hydrodynamics were studied in \cite{Schaefer:2010dv}.
This work shows that quantitative estimates of the shear 
viscosity have to take into account the effects of reheating.

Acknowledgments: This work was supported in parts by the US 
Department of Energy grant DE-FG02-03ER41260. We are grateful
to John Thomas for many useful discussions, and to Jiunn-Wei Chen
for pointing out an error in an earlier version of this 
contribution.


\end{document}